\documentclass[pra,aps,longbibliography,twocolumn,showpacs,superscriptaddress,floatfix,footnoteinbib]{revtex4-1}
\usepackage[latin1]{inputenc}

\usepackage{amssymb}   
\usepackage{amsfonts}
\usepackage{amsfonts}
\usepackage{amsmath}
\usepackage{graphicx}
\usepackage{bm}
\usepackage[dvipsnames]{xcolor}
\usepackage{multirow}
\usepackage[colorlinks, linkcolor=OrangeRed,citecolor=RoyalBlue,urlcolor=RoyalBlue]{hyperref}
\usepackage[all]{hypcap} 

\newcommand{\Le}{\left}
\newcommand{\Ri}{\right}
\newcommand{\nn}{\nonumber}
\newcommand{\f}{\frac}

\newcommand{\eq}[1]{\begin{align}#1\end{align}}

\begin{document}

\title{One-particle density matrix characterization of many-body localization}

\author{Soumya Bera}
\affiliation{Max-Planck-Institut f\"ur Physik komplexer Systeme, 01187 Dresden, Germany}
\author{Thomas Martynec}
\affiliation{Institut f\"ur Theoretische Physik, Hardenbergstr. 36,
Technische Universit\"at Berlin, D-10623 Berlin, Germany}
\author{Henning Schomerus}
\affiliation{Department of Physics, Lancaster University, LA1 4YB Lancaster, United Kingdom}
\affiliation{Max-Planck-Institut f\"ur Physik komplexer Systeme, 01187 Dresden, Germany}
\author{Fabian Heidrich-Meisner}
\affiliation{Department of Physics and Arnold Sommerfeld Center for Theoretical Physics,
Ludwig-Maximilians-Universit\"at M\"unchen, 80333 M\"unchen, Germany}
\affiliation{Kavli Institute for Theoretical Physics, University of California, Santa Barbara CA 93106, USA}
\author{Jens H.\ Bardarson}
\affiliation{Max-Planck-Institut f\"ur Physik komplexer Systeme, 01187 Dresden, Germany}
\affiliation{Department of Theoretical Physics, KTH Royal Institute of Technology, Stockholm, SE-106 91 Sweden}
\date{\today}
\begin{abstract}
We study interacting fermions in one dimension subject to random, uncorrelated onsite disorder, a paradigmatic model of many-body localization (MBL).
This model realizes an interaction-driven quantum phase transition between an ergodic and a many-body localized phase, with the transition occurring in the many-body eigenstates.
We propose a single-particle framework to characterize these phases by the eigenstates (the natural orbitals) and the eigenvalues (the occupation spectrum) of the one-particle density matrix (OPDM) in individual many-body eigenstates. 
As a main result, we find that the natural orbitals are localized in the MBL phase, but delocalized in the ergodic phase. 
This qualitative change in these single-particle states is a many-body effect, since without interactions the single-particle energy eigenstates are all localized.
The occupation spectrum in the ergodic phase is thermal in agreement with the eigenstate thermalization hypothesis, while in the MBL phase
the occupations preserve a discontinuity at an emergent Fermi edge. 
This suggests that the MBL eigenstates are weakly dressed Slater determinants, with the eigenstates of the underlying Anderson problem as reference states.
We discuss the statistical properties of the natural orbitals and of the occupation spectrum in the two phases and as the transition is approached.
Our results are consistent with the existing picture of emergent integrability and localized integrals of motion, or quasiparticles, in the MBL phase.
We emphasize the close analogy of the MBL phase to a zero-temperature Fermi liquid: in the studied model, the MBL phase is adiabatically connected to the Anderson insulator and the occupation-spectrum discontinuity directly indicates the presence of quasiparticles localized in real space.
Finally, we show that the same picture emerges for interacting fermions in the presence of an experimentally-relevant bichromatic lattice and thereby demonstrate that our findings are not limited to a specific model.
\end{abstract}
\maketitle

\section{Introduction}

The perturbative analysis of interacting lattice electrons in the presence of disorder \cite{Basko2006,Gornyi2005} and subsequent numerical and analytical studies \cite{Oganesyan2007,Znidaric2008,Pal2010,Bardarson2012, Bauer2013,Vosk2013, Serbyn2013, Serbyn2013a,Iyer2013,Kjaell2014,Huse2014,Ros:2015ib,Luitz2015,Bera2015,Imbrie2016,Imbrie2016a} of primarily one-dimensional lattice models of fermions with short-range interactions and the related XXZ spin-chain with random field, have firmly established the existence of  the many-body localized (MBL) phase, a perfectly insulating state of interacting particles that persists at any finite energy density.
The insulating behavior manifests itself via a vanishing charge and thermal conductivity \cite{Basko2006,Gornyi2005}.
Depending on the strength of disorder, the MBL phase can transition into an ergodic phase as energy density increases.
This general picture implies the existence of a mobility edge, which in systems with fully localized single-particle states arises as a many-body effect. 
The existence of the mobility edge is supported by numerical investigations \cite{Kjaell2014,Luitz2015,Bera2015,BarLev2015,Naldesi2016}, but has been questioned in Ref.~\cite{DeRoeck2016}.
Many-body localization in systems with a single-particle mobility edge or a fully delocalized single-particle spectrum were studied in Refs.~\onlinecite{Li2015,Modak2015} and \onlinecite{BarLev2016,Sierant2016,Li2016}, respectively.

The precise nature of the transition between the MBL and the ergodic, delocalized phase is still under scrutiny and not well understood (see the discussion in the reviews~\onlinecite{Nandkishore2015,Altman2015}, and, in particular, Refs.~\onlinecite{Potter2015,Vosk2015,Serbyn:2015hr,Zhang:2016kb,Khemani:2016voa,Serbyn:2016wn}). 
The delocalized phase itself has interesting transport properties, with most studies indicating subdiffusive transport in the vicinity of the transition \cite{Agarwal2015,Gopalakrishnan2015,BarLev2015,Luitz:2016ez}, at least for the one-dimensional models that are amenable to numerical studies (some studies have challenged this picture  \cite{Steinigeweg2016,Barisic2016,Bera:2016}).
An interesting aspect of the MBL transition, never observed in any other type of phase transition, is the difference in entanglement scaling in the two phases: the ergodic phase exhibits the usual volume law scaling consistent with thermal behavior and the validity of the eigenstate thermalization hypothesis \cite{Deutsch1991, Srednicki1994,Prosen1999,Rigol2008}, while in the MBL phase all many-body eigenstates exhibit an area law \cite{Bauer2013,Kjaell2014,Friesdorf2015}.

The many-body localization transition takes place in closed isolated systems. 
In solid-state systems it is hard to disconnect the electronic system from the phonon bath provided by the lattice, though it has been suggested that a finite-temperature insulator is obtained in high magnetic field insulating oxides that directly transition into superconductors~\cite{Ovadia:2015ix}.
Ultra-cold quantum gases or systems of trapped ions, however, realize a closed quantum system to a good approximation and therefore, several experiments have investigated the effects of disorder in the presence of interactions at elevated energy densities \cite{Schreiber2015,Kondov2015,Bordia2016,Choi2016,Smith2016,Bordia2016a}.
 A series of optical-lattice experiments  with fermions in 1D \cite{Schreiber2015} and strongly interacting bosons in 2D \cite{Choi2016} provided evidence that MBL is realized and that the transition to an ergodic phase (taking into account system sizes and accessible time scales) can be probed.
Similar conclusions were reached in trapped ion experiments that realize a spin-1/2 chain, albeit with a smaller number of degrees of freedom
\cite{Smith2016}.

The MBL phase is adiabatically connected to the Anderson insulator as interaction strength decreases, at least for the paradigmatic model of 1D spinless fermions with nearest-neighbor interactions exposed to onsite disorder \cite{Oganesyan2007,Pal2010,BarLev2015,Luitz2015,Bera2015}. 
Both phases share many features: they fail to thermalize \cite{Basko2006,Oganesyan2007,Pal2010}, there is an area law of entanglement entropy in all many-body eigenstate \cite{Bauer2013,Kjaell2014}, and there are (quasi-)local conserved quantities \cite{Vosk2013,Serbyn2013,Huse2014,Ros:2015ib,Imbrie2016,Imbrie2016a,Imbrie:2016uj}.
The MBL phase is usually referred to as the generic state, given that in nature interactions are never exactly zero.
Additional phase transitions, in which eigenstates exhibit topological or symmetry-breaking order~\cite{Huse2013,Kjaell2014}, can furthermore occur inside the MBL phase and the possibility of MBL states not adiabatically connected to an Anderson insulator was discussed in Ref.~\cite{Parameswaran2016}.
The Anderson insulator and the MBL state nevertheless differ in several important respects, 
including, for example, the logarithmic increase of entanglement in global quantum quenches \cite{Znidaric2008,Bardarson2012} and  dynamical properties \cite{Agarwal2015,Gopalakrishnan2015}, such as the optical conductivity.

The observation of a logarithmic entanglement growth in global quantum quenches led to the introduction of local 
integrals of motion in the MBL phase~\cite{Vosk2013,Serbyn2013,Huse2014,Ros:2015ib,Imbrie2016,Imbrie2016a,Imbrie:2016uj} (these are often referred to as l-bits but in this paper, for reasons to become clear, we will from here on refer to them as quasiparticles).
In a clean system, a linear spreading of entanglement \cite{DeChiara2006,Calabrese2007} is believed to result from a propagation by contact \cite{Kim2013}, i.e., the fact that degrees of freedom that are spatially close to each other have a direct (off-diagonal) tunnel
coupling. 
For the MBL phase in the regime where all many-body eigenstates are localized, the following  description in terms of an extensive set of conserved quantities has been widely accepted \cite{Vosk2013,Serbyn2013,Huse2014}:
there is an effective Hamiltonian expressed in terms of $L$ mutually commuting quasi-local conserved quantities $ \hat n^{(\rm qp)}_{i}$, with $L$ the number of lattice sites, that are coupled only via products,
\begin{equation}
H = \sum_i  \epsilon_i \hat n_i^{(\rm qp)} + \sum_{ij} J_{ij} \hat n_i^{(\rm qp)} \hat n_j^{(\rm qp)} + \dots\,,
\label{eq:hlbit}
\end{equation}
where the coupling constants $J_{ij}$ decay exponentially with distance between $\hat {n}^{(\rm qp)}_i$ and $\hat{n}^{(\rm qp)}_j$. 
This type of coupling causes a dephasing of spatially separated quasiparticles $\hat{n}^{(\rm qp)}_i$, explaining the logarithmic increase of entanglement as a function of time in quenches from a product state \cite{Serbyn2013a,Nanduri2014}.
The form of the Hamiltonian $\tilde H$ suggests that the MBL phase is integrable in the sense that  $L$ many quasiparticle density operators $\hat n_{i}^{(\rm qp)}$ exist with eigenvalues $0,1$.
Fixing the eigenvalues of all $\hat n^{(\rm qp)}_i$ uniquely specifies a many-body eigenstate in the MBL phase. 
This picture is consistent with the observation of a Poissonian level spacing distribution in the 
MBL phase, the failure of ETH and the vanishing of dc transport coefficients \cite{Nandkishore2015,Altman2015,Imbrie2016a,Imbrie:2016uj}.

For the purpose of this work, let us discuss in more detail the structure of the local integrals of motion.
For the density  operators  $\hat n_i^{(\rm qp)} = c^{(\rm qp)\dagger}_i c^{(\rm qp)}_i$ we can make an ansatz for its expansion in terms of the original fermions
$c_i$ where we ignore spin degrees of freedom (alternatively, one can start from the eigenbasis of the Anderson problem) \cite{Ros:2015ib,Rademaker2016,Imbrie2016a}:
\begin{eqnarray}
\hat n^{(\rm qp)}_{i} &=& \sum_j A_j^{1,(i)} \hat n_j + \sum_{j\not= k}  B_{j,k}^{1,(i)} c^\dagger_j c_k \nonumber  \\
	     &= & \sum_{j,k,l,m} C^{2,(i)}_{j,k,l,m} c^{\dagger}_j c^{\dagger}_k c_m  c_{l} +\dots
\label{eq:lbits}
\end{eqnarray}
Here it is important that $A_j^{1,(i)}$ has its largest weight for $i=j$ and that $A_j^{1,(i)} \to 1$ as interactions and hopping are turned off, while all other coefficients vanish in that limit. 
Thus, the $c^{(\rm qp)}_{i}$ describe a quasiparticle that consists of a particle that is localized in real space over a  distance $\xi^{(1)}_{\rm qp}$ ($A_j^{1,(i)}$ decays exponentially as $\exp(-|i-j|/\xi^{(1)}_{\rm qp})$), dressed with  particle-hole pairs.
The existence of such localized quasiparticles in the MBL phase can in fact be concluded from the  work by Basko, Aleiner and Altshuler \cite{Basko2006}, where the divergence of the quasiparticle lifetime in the localized phase was demonstrated.  
Thus, we can equivalently talk about either l-bits or (localized) quasiparticles in characterizing the MBL phase.
In the following, we will use the latter to refer to the quasi-local integrals of motion.
The actual construction of local integrals of motion, or the related quantum circuit that could generate them, for a given model is an actively studied problem \cite{Ros:2015ib,Rademaker2016,Inglis2016,Chandran2015,Chandran:2015bo,Kim:2014wl,Monthus:2016im,Khemani2016,Pollmann2016,Monthus:2016go,Pekker:2014wia,Yu:2015wya,OBrien:2016wwa,Pekker:2016vna,He:2016wya,Wahl:2016uv}.

The existence of localized quasiparticles in the MBL phase as well as the form of Eq.~\eqref{eq:hlbit} suggest a close analogy of the MBL phase to a (zero-temperature) Fermi liquid, where in the latter, the quasiparticles are physical electrons dressed with particle-hole pairs.
Besides the physical origin of the quasiparticles (restricted phase space versus localization), a key difference is that the relevant quantum number of the quasiparticles in a Fermi-liquid is momentum, while in the MBL phase, the quasiparticle can be labelled by a real-space index.
This picture suggests that the MBL phase relates to the Anderson insulator in a similar way as the Fermi-liquid relates to
the free Fermi gas. 
The important difference that we will emphasize during our discussion is the fact that the MBL is a robust
state at any energy density, while the Fermi liquid is an asymptotic description of a real interacting Fermi gas in the limit
of low energies $E \to E_0$ only, where $E_0$ is the ground-state energy.

A hallmark feature of a Fermi liquid is the finite discontinuity $0<\Delta n_k<1$ at the Fermi energy in the momentum distribution function $n_k = \langle c_k^\dagger c_k\rangle$, where the discontinuity is related to the quasiparticle weight $Z_k$.
In our previous work \cite{Bera2015}, we demonstrated that a similar discontinuity is also present in the occupation spectrum of the one-particle density matrix (OPDM) in the MBL phase (see also earlier analyses of the OPDM in 1D systems of hard-core bosons in the presence of disorder \cite{Nessi2011,Gramsch2012}).
Specifically, we computed the OPDM $\rho^{(1)}_{ij}= \langle n | c^{\dagger}_i c_j| n \rangle $ in individual many-body eigenstates $|n\rangle$.
The diagonalization of the OPDM  yields a basis set in the single-particle subspace and their occupations:
\begin{equation}
\rho^{(1)} | \phi_{\alpha}\rangle = n_{\alpha} | \phi_{\alpha}\rangle\,.
\end{equation}
In the Anderson insulator every many-body eigenstate is a Slater determinant and the occupation spectrum  $n_{\alpha}$ (assuming that the $n_\alpha$ are ordered $n_1 \geq n_2 \geq \dots  \geq n_L$) is a step function $n_{\alpha} =1 $ for $1\leq \alpha\leq N$, where $N$ is the particle number, and zero otherwise. 
This is true irrespective of the disorder strength.
In the MBL phase and with finite hopping and interactions, by contrast, the occupations start to deviate from zero or one in the vicinity of $\alpha=N$.  
Nevertheless, our earlier work \cite{Bera2015} indicated that the spectrum  still exhibits a discontinuity $\Delta n = n_{N+1} -n_N <1$.
The existence of such a discontinuity is clearly incompatible with ergodic and thermal behavior since $\Delta n \to 0 $ is expected in the delocalized phase as system size increases \cite{Bera2015}.
Based on the analysis of the eigenvalues of the OPDM,  we therefore identified 
another means to distinguish the MBL from the ergodic phase.

Apart from the occupation spectrum, the eigenstates of the OPDM, commonly referred to as natural orbitals, bear interesting information as well:  they are typically localized in the MBL phase and delocalized in the ergodic phase. 
Moreover, they exhibit an exponential envelope in the MBL phase with a natural-orbital localization length that we refer to as $\xi_{\rm NO}$.
It is natural to use this quantity as a measure of the effective single-particle localization length in MBL phase.
These properties suggest that  these natural orbitals can be analyzed in much the same way as single-particle eigenstates in the Anderson insulator, for example using appropriately defined inverse participation ratios.

In this work we give a comprehensive overview of the properties of the eigenstates (the natural orbitals) and eigenvalues (the occupation spectrum) of the OPDM in a generic interacting 1D system in the presence of uncorrelated diagonal disorder. 
First, as a prelude to the analysis, we discuss the connection between the local integrals of motion and the eigenstates of the OPDM on a qualitative level.
Second, we analyze the natural orbitals  and discuss their dependence on disorder strength, energy density and interaction strength.
Third, we discuss the properties of the occupation spectrum, focussing on the existence of the discontinuity $\Delta n$ in the MBL phase and its disappearance in the  ergodic phase. 
We conclude the discussion of the natural orbitals by arguing that they provide an improved approximation to the quasiparticles compared to other single-particle basis sets.
Finally, we apply our methodology to a system of interacting spinless fermions in the Aubry-Andr{\'e} model,  
showing that 
properties of the OPDM capture the many-body localization physics in this model as well.
We conclude with a summary of our results and an outlook on open problems, with a focus on those cases
in which we believe the analysis of the OPDM could be useful.

\section{Localized conserved quantities and OPDM eigenstates}
\label{sec:lbits}

An important concept in the theory of MBL is that of local conserved quantities~\cite{Vosk2013,Serbyn2013,Huse2014,Ros:2015ib,Imbrie2016,Imbrie2016a,Imbrie:2016uj}. 
We outline here how the eigenstates of the OPDM are related to these objects, which will serve as the basis for the rest of the paper.
A useful starting point is to diagonalize the noninteracting problem, including all one-body potentials and thus also disorder:
\begin{equation}
H_0 = \sum_{\mu} \epsilon_\mu  \hat n_\mu, \label{eq:anderson}
\end{equation}
with $\hat n_\mu = c^\dagger_\mu c_\mu$ the density operator.
We reserve the indices $\mu,\nu,\eta$ for this Anderson single-particle eigenbasis.
For this single-particle problem, the density operators $\hat{n}_\mu$, $\mu = 1,\ldots,L$, comprise, by construction, a set of local mutually commuting conserved quantities.
All the many-body eigenstates are product states of the corresponding creation operators: $|\{n_\mu\}\rangle = \prod_\mu (c^\dagger_\mu)^{n_\mu}|0\rangle$, where $n_\mu \in \{0,1\}$, the eigenvalues of $\hat{n}_\mu$, determine the occupation of a given single-particle state in the many-body state.

In the presence of interactions the conserved quantities are similarly density operators of quasiparticles, which are dressed versions of the single-particle Anderson orbitals.
Explicit algebraic constructions of the integrals of motion \cite{Rademaker2016,Ros:2015ib}, which we here denote as $\hat n_{\mu}^{(\rm qp)}$ to stress the quasiparticle interpretation, result in 
\begin{equation}
 {\hat n_{\mu}^{(\rm qp)}} = \hat n_{\mu} + \sum_{(\nu\nu^\prime) \not= (\eta \eta^\prime)} B^{\mu}_{\nu\nu^\prime\eta\eta^\prime } c^\dagger_\nu  c^\dagger_{\nu^\prime}  
c_{\eta} c_{\eta^\prime}+ \dots \,.
\label{eq:lbits2}
\end{equation}
This particular form ensures that $\lbrack H ,  {\hat n_\mu^{(\rm qp)}} \rbrack=0$ in each $N$-particle subspace, including $N=1$, which requires all off-diagonal quadratic terms to be zero.
The choice of only single diagonal term $\hat n_\mu$ on the right hand side with a prefactor of one makes the identifications of quantum numbers $\mu$ explicit \cite{Ros:2015ib,Rademaker2016}.
Expressing the integrals of motion like this in the Anderson eigenbasis makes it transparent that a  set of $L$ many of them exists for a given disorder realization and by construction, ${\hat n_{\mu}^{(\rm qp)}} \to \hat n_{\mu}$ as interaction strength is sent to zero.
It is further essential in order for the quasiparticle operator $\hat{n}^{(\rm qp)}_{\mu}$ to generally have only eigenvalues 0 and 1, required for the quasiparticle interpretation.
In this case one can introduce the creation and annihilation operators for quasiparticles via $\hat{n}^{\rm (qp)}_\mu = c^{({\rm qp})\dagger}_\mu c^{\rm (qp)}_\mu$. 
Since the Hamiltonian remains diagonal in the $\hat{n}^{(\rm qp)}_\mu$, see the explicit form of the Hamiltonian in Eq.~\eqref{eq:hlbit}, the many-body eigenstates are product states of the quasiparticles: $|n\rangle = |\{n^{\rm (qp)}_\mu\}\rangle = \prod_\mu (c^{(\rm qp)\dagger}_\mu)^{n^{\rm (qp)}_\mu}|0\rangle$.
Here, as in the noninteracting case, $n^{\rm (qp)}_\mu \in \{0,1\}$ are the eigenvalues of $\hat{n}^{\rm (qp)}_\mu$ and give the occupations of a quasiparticle in a given eigenstate.

For each many-body eigenstate $|n\rangle$, the diagonalization of the OPDM by the unitary transformation $U_{\alpha i}^{(n)}$ defines a new complete set of single particle operators via
\begin{equation}
	c_{\alpha}^\dagger = \sum_i U_{\alpha i}^{(n)} c_i^\dagger = \sum_\mu \tilde{U}^{(n)}_{\alpha\mu} c_\mu^\dagger\,.
\label{eq:ualphai}
\end{equation}
The column entries of $U^{(n)}$ are the natural orbitals $\phi_{\alpha}(i)$ and $\tilde{U}$ gives their projection onto the Anderson orbitals. 
The creation operator $c^\dagger_\alpha$ therefore creates a particle in the natural orbital $\phi_\alpha$.
The indices $\alpha,\beta,\gamma,\delta$ are reserved for the OPDM eigenbasis.
Since these states are obtained in a given many-body eigenstate, the operators $c_{\alpha}^\dagger$ are in general state dependent.
We emphasize that therefore, any quantity below that has an index $\alpha,\beta,\gamma,\delta$ depends in general on the state $|n\rangle$.

By inserting this transformation into the quasiparticle expansion Eq.~\eqref{eq:lbits2}, we obtain the representation of the integrals of motion in terms of the creation and annihilation operators of the natural orbitals 
\begin{eqnarray}
        \hat{{n}}_\mu^{(\rm qp)} &=& \sum_{\alpha} |\tilde{U}_{\alpha\mu}|^2 \hat{n}_\alpha + \sum_{\alpha \neq \beta} \tilde{U}^*_{\alpha\mu}\tilde{U}_{\beta\mu}c^\dagger_\alpha c_\beta \nonumber \\
	&& + \sum_{\alpha\beta\gamma\delta} \tilde{B}^{\mu}_{\alpha\beta\gamma\delta} c^\dagger_\alpha c^\dagger_{\beta}c_{\gamma}c_{\delta} + \ldots,
\label{eq:lbit_exp_opdm}
\end{eqnarray}
where $\hat{n}_\alpha = c^\dagger_\alpha c_\alpha$ and $\tilde{B}$ is obtained from the appropriate contraction of $\tilde{U}$ with $B$.
This basis change generates off diagonal quadratic terms, but the natural orbital basis is special in that these terms vanish when this expression is evaluated in the state $|n\rangle$.
This results directly from their definition
\begin{equation}
        \langle n |c^\dagger_\alpha c_\beta|n\rangle = \delta_{\alpha\beta}n_\alpha,
\end{equation}
with $n_\alpha$ the OPDM occupations.
To leading order, we therefore have the following relation between the quasiparticle occupations and the OPDM occupations
\begin{equation}
        {n}^{(\rm qp)}_\mu =  \sum_\alpha|\tilde{U}_{\alpha\mu}|^2 n_\alpha + \ldots,
\label{eq:qpcoeffs}
\end{equation}
where ${n}^{(\rm qp)}_\mu = 0$ or $1$ give the occupation of the quasiparticles in the eigenstate $|n\rangle$. 
This suggests that as long as $|\tilde{U}_{\alpha\mu}|^2$ is strongly peaked at a particular value $\alpha = \alpha(\mu)$ the occupations of the OPDPM are in a one-to-one correspondence with the occupations of the quasiparticles. 
This is corroborated by our numerical analysis presented in Sec.~\ref{sec:nos}, where we show that the natural orbitals generally have a well defined localization center and exponential tails.
Consequently, since the Anderson orbitals are also exponentially localized, $|\tilde{U}_{\alpha\mu}|^2$ is a sharply peaked function. 
In this case, a value of ${n}^{(\rm qp)}_\mu = 0$ or $1$ implies that $n_{\alpha(\mu)}$ is also close to $0$ or $1$, and the ideal step function structure of the quasiparticle occupations is reflected in a similar discontinuous structure in the OPDM occupations.
This is indeed what we find, see for example Fig.~\ref{fig:nalpha}.

\section{Model}
\label{sec:model}

The model we consider consists of spinless fermions on a 1D lattice with $L$ sites, hopping constant $t$ and nearest and next nearest density-density interaction with strength $V$ and $V'$, given by the Hamiltonian
\eq{
H = t\sum_{i=1}^L &\Le\lbrack -\frac{1}{2} (c^{\dagger}_i c_{i+1} + c_{i+1}^\dagger c_i) +  \epsilon_i \Le(\hat n_i-\f{1}{2}\Ri) \Ri.  \nn \\
 & \Le. + V \Le(\hat n_i-\f{1}{2}\Ri) \Le(\hat n_{i+1}-\f{1}{2}\Ri) \Ri. \nn \\
 &  \Le. + V' \Le(\hat n_i-\f{1}{2}\Ri) \Le(\hat n_{i+2}-\f{1}{2}\Ri)  \Ri\rbrack\,.
\label{eq:ham}
}
Here $c_i^{\dagger}$ is the fermion creation operator on site $i=1,2,\ldots,L$, with the periodic boundary conditions $c^\dagger_{L+1}=c^\dagger_1$, and $\epsilon_i \in \lbrack -W, W \rbrack$ is a random  scalar onsite potential of disorder strength $W$.
We obtain eigenstates at a fixed filling $N = L/2$ by exact diagonalization, and select the energy of eigenstates to match a target
energy density $\varepsilon = 2(E - E_\text{min})/(E_\text{max}-E_\text{min})$, where $E_\text{max}$ and $E_\text{min}$
are the maximum and minimum energy for a given disorder realization.
After obtaining $E_{\rm min}$ and $E_{\rm max}$, we use the shift-invert mode of the standard
ARPACK diagonalization routine~\cite{arpack} to calculate few~(typically 16) eigenstates of the many-body
Hamiltonian~\eqref{eq:ham}.
We show results for both $V'=0$ and $V'>0$. The next-nearest-neighbor interaction is introduced to break the integrability of the
clean model with $W=0$. However, sufficiently far away from the clean case, we did not observe any influence of a nonzero $V'$ on the results, in line
with the suggestion that the MBL phase is robust against microscopic details of the Hamiltonian.
We study this system using exact diagonalization at finite sizes $L=10,12,14$ ($10^5$ disorder realizations), $L=16$ ($10^3$ realizations) and $L=18$ ($500$ realizations), unless stated otherwise. 

\section{Properties of the natural orbitals}
\label{sec:nos}

\subsection{Typical natural orbitals}
\label{sec:nos_examples}

\begin{figure}[tb]
\includegraphics[width=\columnwidth]{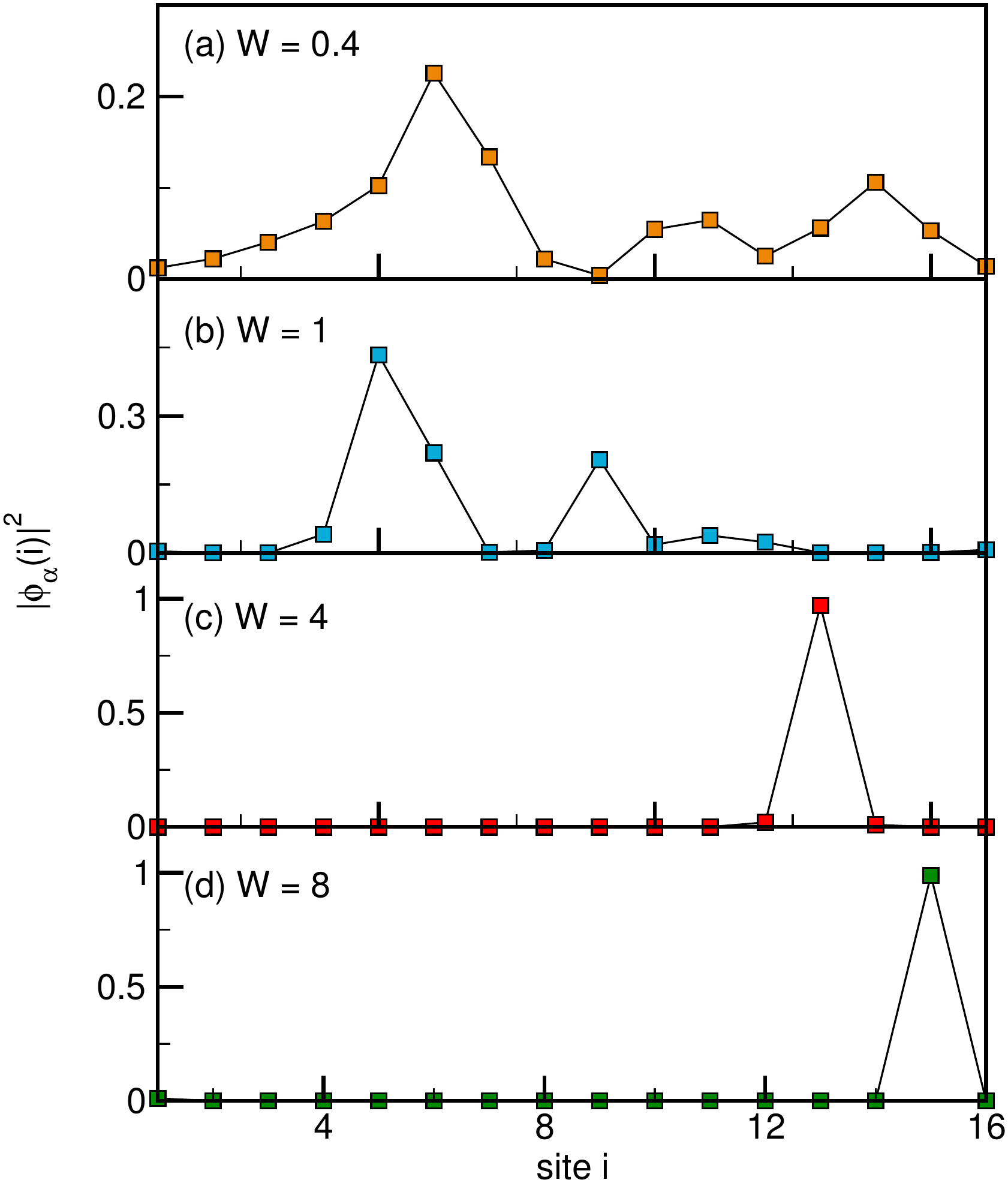}
\caption{(Color online)
Spatial profile of typical natural orbitals in the (a),(b)  ergodic phase ($W=0.4,1$) and (c),(d) the MBL phase ($W=4,8$).
$L=16$, $V=1$, $V'=0.2$, $\varepsilon=1$.}\label{fig:nos_example}
\end{figure}

Figure~\ref{fig:nos_example} shows the spatial profile $|\phi_{\alpha,i}|^2$  of typical natural orbitals $\phi_{\alpha}(i)=\langle i | \phi_\alpha\rangle$ at $V=1, V'=0.2, \varepsilon=1$ and different  disorder strengths. 
In the ergodic phase,  $\mathcal{O}(L)$ many sites carry a nonzero weight. In the MBL phase, however, the vast majority of the natural orbitals 
 possess an easily recognizable single maximum with a significantly larger weight than on all other sites;  we associate this site with the localization center of the corresponding quasiparticle. 
Thus, we summarize one of the main qualitative results of Ref.~\cite{Bera2015} as follows: while all single-particle eigenstates are localized in our model, interactions lead to a delocalization of natural orbitals in the ergodic phase. 
In the MBL phase the natural orbitals retain the main features of Anderson eigenstates, i.e., a well-defined localization center and exponential tails (see the discussion below).

These statements are substantiated by an analysis of the statistical properties of the natural orbitals over the disorder ensemble. 
One measure for the localization or delocalization of a single-particle state is the inverse participation ratio
\begin{equation}
\mbox{IPR} = \frac{1}{N} \sum_{\alpha=1}^L n_{\alpha} \sum_{i=1}^L |\phi_\alpha(i)|^4 \,.
\end{equation}
In the ergodic case, one expects $\phi_\alpha(i) \sim 1/\sqrt{L}$ and hence IPR~$\sim 1/L$, while in the MBL phase and in the case of 
a single localization center, IPR $\sim 1$. 
We presented results for the so-defined IPR in \cite{Bera2015}, which confirm this expectation: the distribution of IPR values $\mathcal{P}(\mbox{IPR})$ is centered around $1/L$ in the ergodic phase and becomes narrower as $L$ increases. 
In the MBL phase, the distributions are broad, centered around a large value that moves towards $\mbox{IPR} \sim 1$ as $W$ increases, and the distributions are $L$ independent.

\begin{figure}[tb]
\includegraphics[width=\columnwidth]{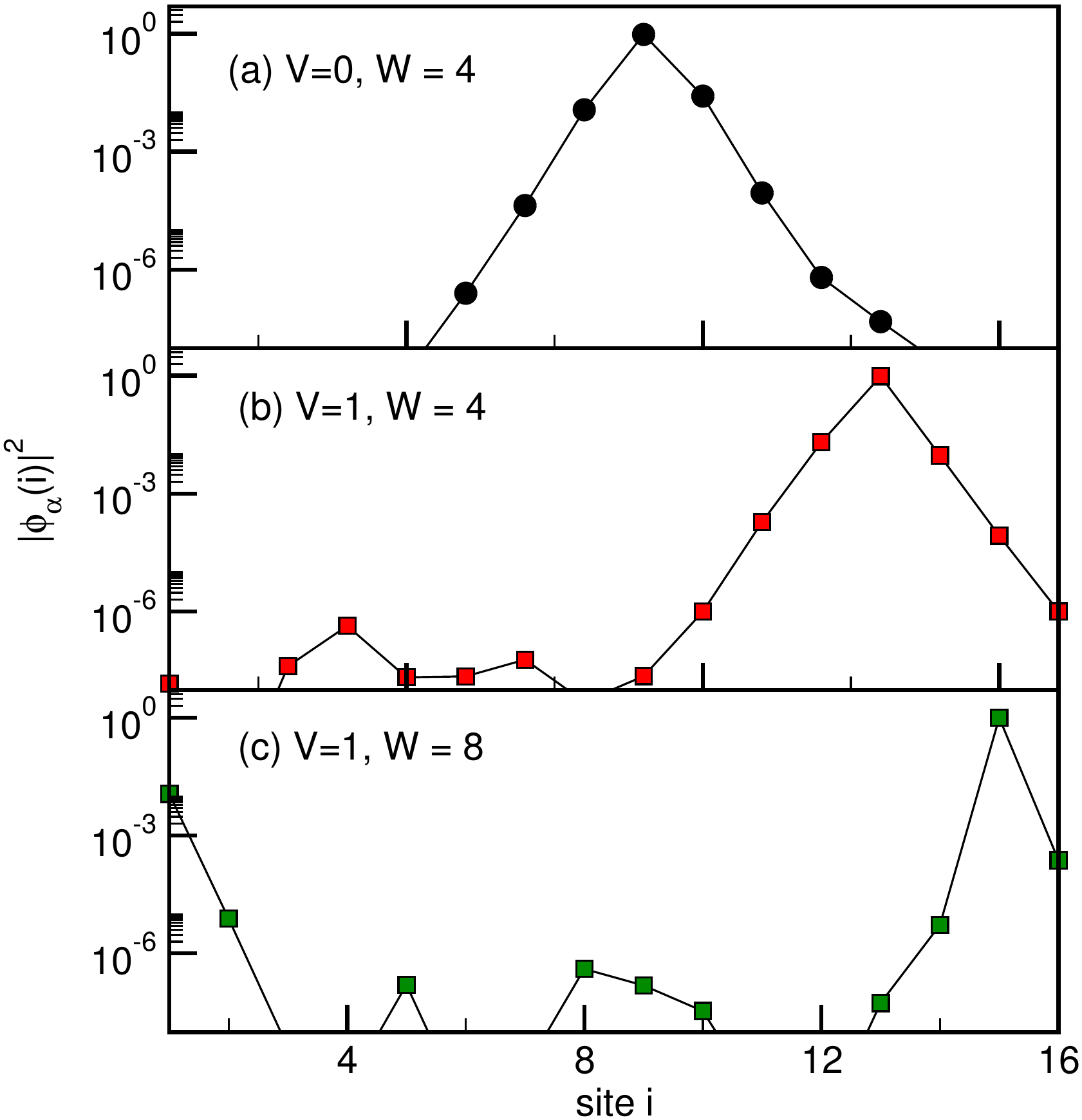}
\caption{(Color online)
Natural orbitals for (a) $V=0$, $W=4$, (b) $V=1, V'=0.2$, $W=4$, $\varepsilon=1$, (c) $V=1, V'=0.2$, $W=4$, $\varepsilon=1$ (both for $L=16$).  The states shown in (a) and (b) are computed for different disorder realizations.}\label{fig:nos_loc}
\end{figure}

Another important feature of the eigenstates of the Anderson problem in one dimension is their exponential tails \cite{Kramer1993} as shown,  for example,  in Fig.~\ref{fig:nos_loc}(a).
These tails directly visualize the (single-particle) localization length in the Anderson problem. 
In Figs.~\ref{fig:nos_loc}(b) and (c), we plot the two natural orbitals from Figs.~\ref{fig:nos_example}(c) and (d) ($\varepsilon=1$) 
which represent states just beyond the transition and deep in the MBL phase, respectively.  
Clearly, these natural orbitals decay exponentially away from the localization center (note the log-lin scale in Fig.~\ref{fig:nos_loc}).
Note that increasing $W$ makes the natural orbitals more narrow (compare Figs.~\ref{fig:nos_loc}(b) and (c)).

In principle, one can therefore extract a  single-particle localization length $\xi_{\rm NO}$ from the natural orbitals, with the expectation that $\xi_{\rm NO} \sim \xi_{\rm qp}^{(1)}$ of the {\it quasiparticles} in the MBL phase. 
However, from such small systems as accessible to exact diagonalization or even the shift-and-invert method \cite{Luitz2015}, this would not reliably work even in the noninteracting case, due to the fluctuations in the single-particle states around their exponential envelope. 
Moreover, on the system size studied, not all natural orbitals exhibit a clear exponential tail in the MBL phase. 
At present, it is unclear whether this is due to the small system sizes or an actual property of the states.
We thus postpone this very interesting question, namely the extraction of the localization length $\xi_{\rm NO}$ 
from the natural orbitals, and the analysis of their statistical properties as well as  
their dependence on disorder strength, to a future study using the DMRG methods that recently have been developed 
specifically for the MBL phase \cite{Khemani2016,Lim2016,Kennes2016,Yu:2015wya, Pollmann2016,Serbyn:2016gv}.
System sizes of $L\sim 100$ should be accessible with these techniques, at least deep in the MBL phase.

One may object that the natural orbitals are not uniquely defined because of possible degeneracies in the eigenspectrum of the OPDM. 
This is certainly true for the noninteracting case without disorder.
In the presence of disorder and interactions,  there are  typically no degeneracies in the occupation spectrum for the system sizes considered here for a {\it single} disorder realization and within the numerical accuracy. We cannot rule out  that on larger systems, true degeneracies in the occupation spectrum will emerge (which would require a more careful
analysis of the natural orbitals). 

%

\subsection{Energy- and $V$-dependence}
For every many-body eigenstate $|n\rangle$, the diagonalization of the OPDM defines a unitary transformation from the 
original real-space basis to a new basis labelled with $\alpha$, as given in Eq.~\eqref{eq:ualphai}.
There are several obvious questions about the properties of the $c_{\alpha}$ and the transformations $U_{\alpha j}$. 
First, can one give some physical interpretation to the index $\alpha$? 
In the clean case, since the OPDM respects lattice translational symmetry, $\alpha$ can be identified with quasimomentum $k$
and hence there is no information in the natural orbitals as one varies interaction strength; they are always plane waves.
In the MBL phase, the discussion from Sec.~\ref{sec:nos_examples} suggests that $\alpha$ can still be thought of as a site index, indicating the localization center of a quasiparticle.
In the ergodic phase, there is no obvious way to interpret the index $\alpha$.

\begin{table}
\begin{tabular}{|r|c|l|l|l|l|}
\hline
\mbox{$i$} & $\epsilon_i$ & \multicolumn{4}{|c|}{$n_{\alpha}$ } \\
\hline
\multicolumn{2}{|c|}{}                       & $V=0$ & $V=0.05$ & $V=0.1$ &   $V=0.25$  \\
\hline
 1	&	5.42402845 	& 0	& 1.00			&	$2.5 \,10^{-7}$	&	1.00 \\		
 2	&	6.40946998	& 1	& 2.48 $10^{-8}$	&	$1.00$	&	1.00 \\
 3	&	-7.14347121	& 1	& 1.00			&	$9.38 \,10^{-8}$	&	1.00 \\
 4	&	3.21053334	& 1	& 1.00			&	$1.00$	&	$5.22 \, 10^{-7}$ \\
 5	&	-4.22123147	& 0	& 1.00			&	$6.44 \, 10^{-7}$	&	1.00 \\
 6	&	-4.04111029	& 1	& 5.49 $10^{-9}$	& 	$7.03 \, 10^{-7}$	&	1.00 \\
 7	&	-5.21082024	& 0	& 3.07 $10^{-8}$	&	$1.00$	&	$1.77 \, 10^{-6}$\\
 8	&	-6.94881084 	& 0	& 2.88 $10^{-8}$	&	$1.00$	&	$3.10 \, 10^{-6}$\\
 9 	&	5.87062939 	& 1	& 7.54 $10^{-13}$	&	$1.43 \, 10^{-6}$	&	$7.33 \, 10^{-6}$\\
10 	&	1.31757566	& 0	& 7.38 $10^{-13}$	&	$1.00$	&	$1.00$ \\
11	&	-1.73720442 	& 1	& 1.83 $10^{-8}$	&	$1.00$	&	1.00\\
12	&	-7.12339691 	& 1	& 1.00	&	$1.48 \, 10^{-6}$	&	$5.90 \, 10^{-6}$\\
13	&	7.16553615	&  0	& 1.00	&	$2.38 \, 10^{-7}$	&	$7.38 \, 10^{-7} $ \\
14	&	-0.62631181	& 0	& 1.00	&	$1.00$	&	$2.92 \, 10^{-7}$ \\ \hline
\end{tabular}
\caption{Onsite potentials $\epsilon_i$, occupations $n_\alpha$, and localization centers $i$ for the natural orbitals for $\varepsilon=1, W=8$ but different $V=0,0.05,0.1, 0.25$.}
\label{tab:params_v}
\end{table}

Second, how do the natural orbitals change as either interaction strength $V$ or energy density $\varepsilon$ varies?
To address this point we consider a fixed disorder strength of $W=8$ deep in the MBL phase and {\it one} disorder realization that is also kept fixed as $V$ or $\varepsilon$ are varied.
The values of the onsite potentials $\epsilon_i$ and the occupations $n_\alpha$  are listed in Tab.~\ref{tab:params_v}
and Tab.~\ref{tab:params_e} for fixed $\varepsilon=1$ and fixed $V=1$, respectively. 
The site index listed in the first column is the localization center of  the natural orbital $\phi_\alpha(i)$, for which the $n_\alpha$ is quoted.

\begin{figure}[tb!]
\includegraphics[width=1\columnwidth]{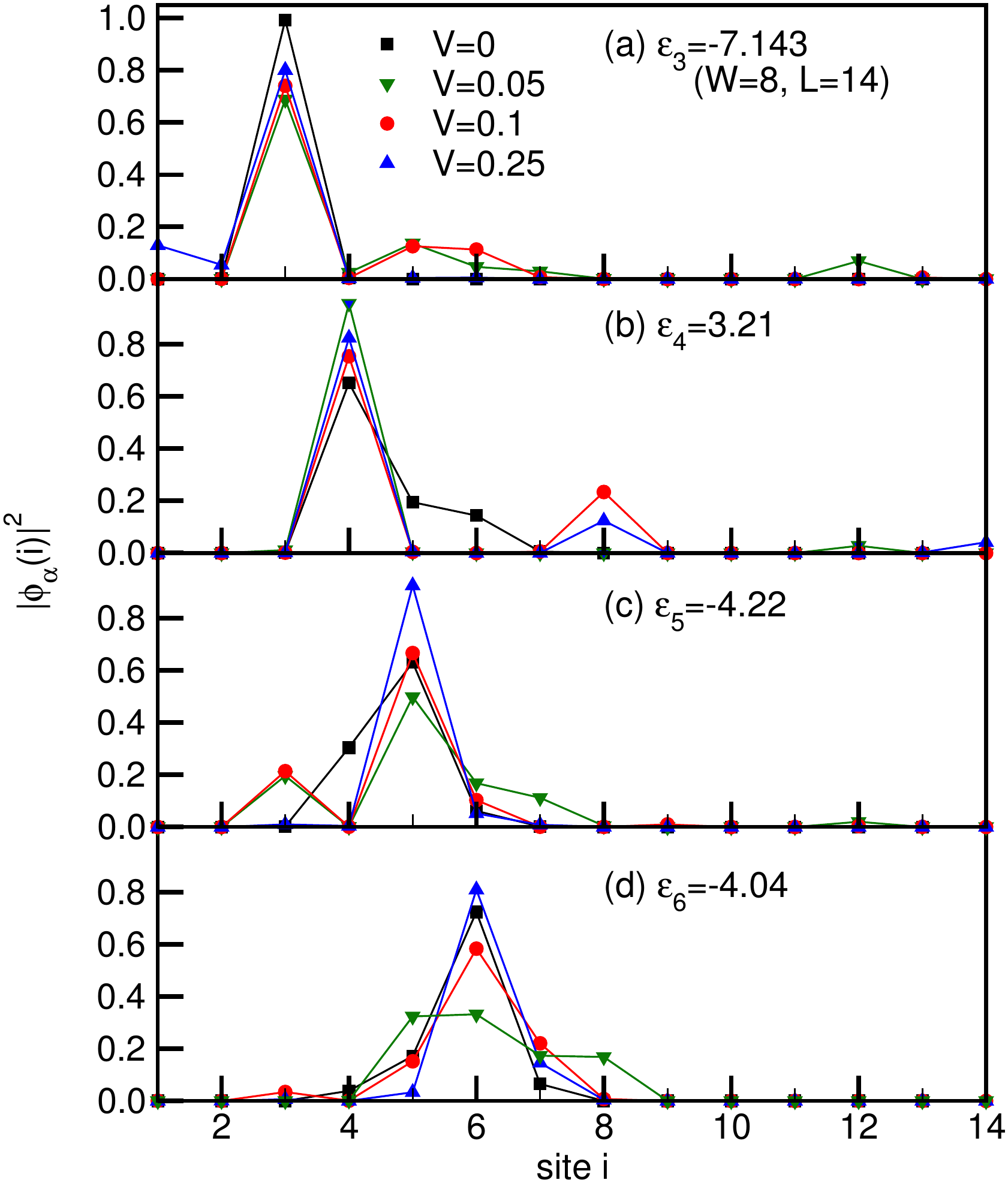}
\caption{(Color online) Natural orbitals for one disorder realization at $W=8$ but different interaction strengths $V=0, 0.05, 0.1, 0.25$ ($V'=0$).
The natural orbitals for the interacting case are computed from a state at energy density $\varepsilon=1$ and we show the natural orbitals
with localization centers at sites (a) $i=3$, (b) $i=4$, (c) $i=5$, (d) $i=6$.
}
\label{fig:singleNOs_v}
\end{figure}

Let us first discuss the $V$-dependence.
Figures~\ref{fig:singleNOs_v}(a)-(d) show four different natural orbitals at $\varepsilon=1$ centered around sites $i=3,4,5,6$,
respectively, for $V=0,0.05,0.1,0.25$.
We select these sites since  $\epsilon_3$ is very low, which should typically localize a particle there at sufficiently
low $\varepsilon$. 
The sites $i=5,6$ have comparable onsite potentials $\epsilon_5\approx \epsilon_6$.

As expected, the Anderson state for $i=3$ has a very simple structure with its largest weight at the localization center. Increasing $V$ changes this (and other
natural orbitals) quickly and in an apparently unsystematic way: the $i=3$ natural orbitals for $V=0.05$ and $V=0.1$ are more spread out while the one for $V=0.25$
looks practically like the corresponding Anderson eigenstate (see Fig.~\ref{fig:singleNOs_v}(a)).
Note that this observation  makes no implication on the $V$-dependence of the localization length $\xi_{\rm NO}$,
since those need to be computed from full distributions from many disorder realizations.

The natural orbitals centered around $i=4,5,6$ are already hybridized in the sense that all three of them have
sizable weight on all three sites (still with a dominant maximum at the localization center). That structure is also not
preserved systematically in the natural orbitals for $V>0$, while still, hybridizations between sites $i=5$ and $i=6$
frequently occur (see, e.g., the results for $V=0.25$).

The occupations listed in Tab.~\ref{tab:params_v} show that whether a state centered around a given site $i$ is occupied or not depends on $V$.
This result is not surprising as fixing energy has no particular meaning in the nonergodic MBL phase. 
Rather, many-body eigenstates in the MBL phase are  specified by selecting $N$ nonzero eigenvalues of 
$n^{\rm (qp)}_i$, all other $L-N$ being zero.
To see the evolution of the natural orbitals from the Anderson eigenstates as interactions are turned on, one 
therefore has to select a set of occupations ${ n^{\rm (qp)}_1, \dots ,n^{\rm (qp)}_L }$ of single-particle eigenstates at $V=0$ for a
disorder realization that is kept fixed.
Then, as interactions are turned on to small but finite values $V>0$, one needs to diagonalize the full $N$-body subspace
and select the one state whose real-space structure has the largest overlap with the non-interacting many-body state.
This can be judged by diagonalizing the OPDM in each many-body eigenstate and by associating each $n_\alpha$ to  the localization
center of the corresponding natural orbital.  This set of natural orbitals is then the one that should evolve adiabatically
from the original set of Anderson eigenstates as interactions are added, while its eigenenergy can evolve through the
spectrum.

\begin{table}
\begin{tabular}{|r|c|l|l|l|l|}
\hline
\mbox{$i$} & $\epsilon_i$ & \multicolumn{4}{|c|}{$n_{\alpha}$ } \\
\hline
\multicolumn{2}{|c|}{}                       & $\varepsilon=0$ & $\varepsilon=1/3$ & $\varepsilon=2/3$ &   $\varepsilon=1$  \\
\hline
 1      &       5.42402845      & 1.72 $10^{-11}$  & 1.88 $10^{-5}$    	&     1.00   	       & 3.76 $10^{-8}$      \\
 2      &       6.40946998      & 1.530 $10^{-8}$  & 1.00 		&     8.75  $10^{-6}$  & 2.89 $10^{-7} $      \\
 3      &       -7.14347121     & 1.00     	   & 1.00               &     1.00 	       & 1.00     \\
 4      &       3.21053334      & 1.531 $10^{-8}$  & 9.32 $10^{-6}$     &     1.01 $10^{-5}$   & 1.00     \\
 5      &       -4.22123147     & 1.00     	   & 1.00 		&     8.18 $10^{-6}$   & 2.00 $10^{-6}$      \\
 6      &       -4.04111029     & 1.00     	   & 1.00       	&     1.00 	       & 1.74 $10^{-6}$       \\
 7      &       -5.21082024     & 1.00     	   & 1.31 $10^{-5}$   	&     1.08 $10^{-6}$   & 9.30 $10^{-7}$      \\
 8      &       -6.94881084     & 1.00     	   & 1.00        	&     2.42 $10^{-5}$   & 1.00       \\
 9      &       5.87062939      & 4.05 $10^{-7}$   & 2.58 $10^{-7}$     &     2.33 $10^{-5}$   & 1.00      \\
10      &       1.31757566      & 7.07 $10^{-7}$   & 5.64 $10^{-9}$     &     1.00             & 1.00     \\
11      &       -1.73720442     & 1.00     	   & 1.44 $10^{-7}$   	&     1.00             & 5.17 $10^{-6}$      \\
12      &       -7.12339691     & 1.00     	   & 1.00  		&     1.00             & 1.00     \\
13      &       7.16553615      & 3.09 $10^{-7}$   & 1.46 $10^{-5}$  	&     5.05 $10^{-6}$   & 1.00      \\
14      &       -0.62631181     & 6.18 $10^{-9}$   & 1.00	 	&     1.00             & 4.37 $10^{-6}$      \\ \hline
\end{tabular}
\caption{Onsite potentials $\epsilon_i$, occupations $n_\alpha$, and localization centers $i$ for the natural orbitals for $V=1, W=8$ but different $\varepsilon=0,1/3,2/3,1$.}
\label{tab:params_e}
\end{table}

\begin{figure}[tb!]
\includegraphics[width=1\columnwidth]{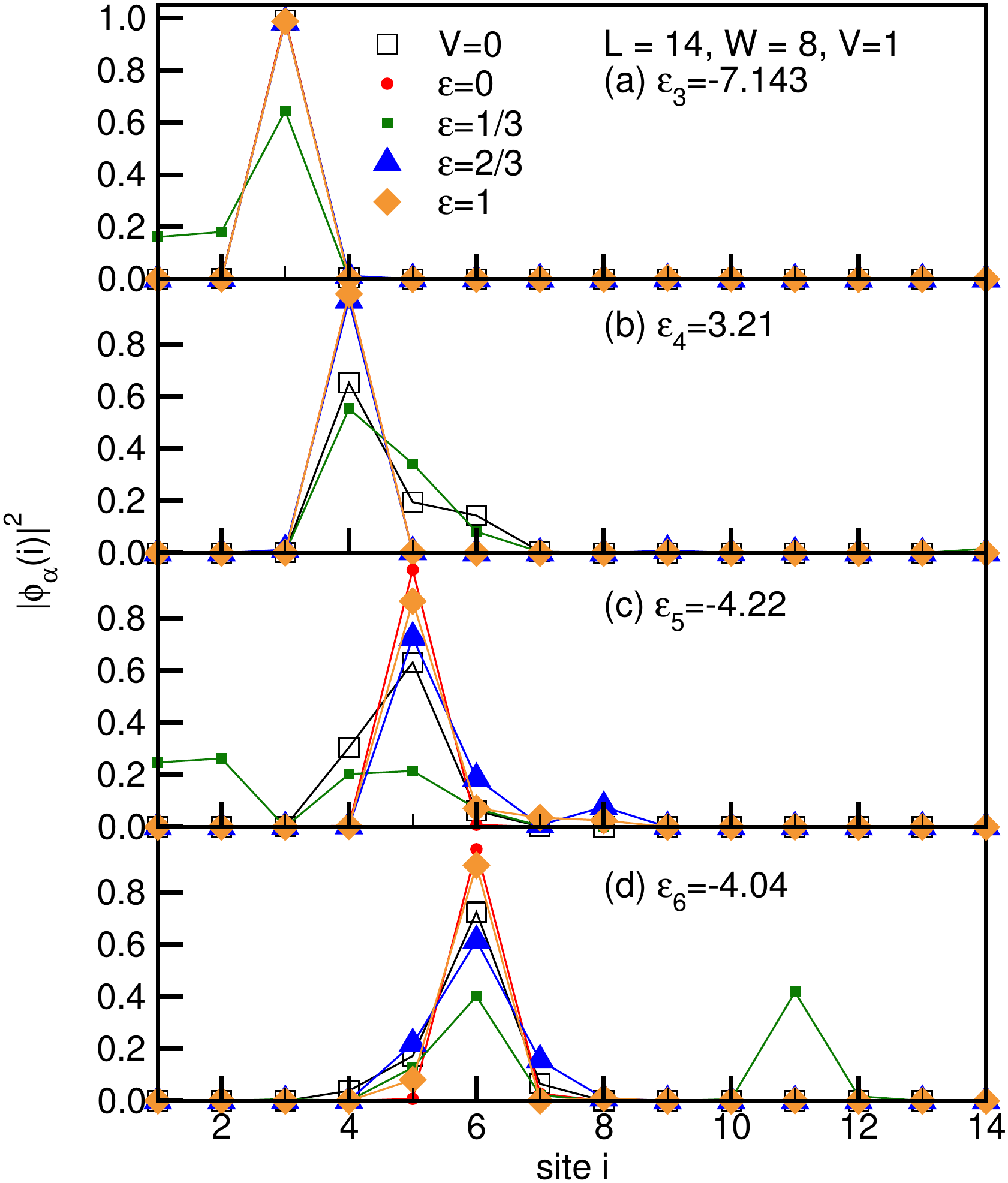}
\caption{(Color online) Natural orbitals for one disorder realization at $W=8$ and $V=1, V'=0$ but different energy densities $\varepsilon=0,1/3,2/3,1$.
We also include the corresponding Anderson eigenstates. The figures shows the natural orbitals for sites (a) $i=3$, (b) $i=4$, (c) $i=5$, (d) $i=6$.
}
\label{fig:singleNOs_e}
\end{figure}

We next turn to the energy dependence of the natural orbitals. We present a set of natural orbitals centered around sites $i=3,4,5,6$
for $V=1$ in Figs.~\ref{fig:singleNOs_e}(a)-(d) (the Anderson states are included for comparison).
In the ground state ($\varepsilon=0$), the OPDM  occupations (see Tab.~\ref{tab:params_e}) follow the naive expectation that
predominantly the sites with the lowest $\epsilon_i$ are occupied. As energy density increases, different sets of $N$-many sites become occupied. 
The natural orbitals turn out to depend on $\varepsilon$ as there is no systematic way in which they have a simple structure (i.e., a large weight on only
one site) or hydridizations (i.e., multiple sites with $0.1 <|\phi_{\alpha}|^2<1$).

The key reason is that energy is {\it not} the relevant quantity to distinguish many-body eigenstates in the MBL phase, rather the defining property
is the local structure as imprinted onto the state by the $N$ occupied quasiparticles. 
As we discussed in Sec.~\ref{sec:lbits}, it is in general expected that the natural orbitals depend on the state from the relationships between 
quasiparticles and the (state-dependent) OPDM eigenstates.


%
\section{Occupation spectrum}

\subsection{Limiting cases: Clean system, Anderson insulator and deep MBL regime}

The other information obtained from the diagonalization of the OPDM is the occupation spectrum $n_{\alpha}$. 
For a clean system with periodic boundary conditions $\alpha =k$ is just quasimomentum and $n_\alpha$
is thus nothing but the quasimomentum distribution function. In an ergodic system (and away from integrability)
we expect $n_\alpha$ to be a thermal distribution \cite{Wright2014}, based on studies of quantum quenches 
in nonintegrable models (see, e.g., \cite{Rigol2008,Sorg2014}).
We will demonstrate that this expectation holds true in Sec.~\ref{sec:nthermal}.

There are two other limiting cases, in which the structure of $n_\alpha$ is obvious. First, consider the 
Anderson model in 1D:
\begin{equation}
H_0 = -\frac{t}{2} \sum_i (c_i^\dagger c_{i+1} + c^\dagger_{i+1}c_i) + t\sum_i \epsilon_i \left(\hat n_i -\frac{1}{2}\right).
\end{equation}
In that case (excluding degeneracies), the diagonalization of the OPDM will just give the single-particle 
eigenstates.
Since a many-body eigenstate of a noninteracting system of fermions is just a Slater determinant, 
after ordering the OPDM eigenvalues  according to their size $n_1 > n_2 > \dots > n_N > \dots > n_L$, we obtain a simple step function with
$n_{\alpha} = 1$ for $1\leq \alpha \leq N$ and zero otherwise.

Deep in the MBL phase where $W \gg 1,V$, we can approximate the Hamiltonian by
\begin{equation}
H \approx   t\sum_i \epsilon_i \left(\hat n_i -\frac{1}{2}\right).
\end{equation}
The OPDM eigenstates are now simply fully localized on individual sites, and in each eigenstate $N$ of them are occupied. 
In that case, the many-body wave-function is a Slater determinant again and $n_\alpha$ is a step function.
Thus, in these two limits, we can think of $\alpha =N$ corresponding to a Fermi-index labeling the last occupied state.
We can define a Fermi-discontinuity $\Delta n$ from 
\begin{equation}
\Delta n = n_{N+1} - n_N\,.
\end{equation}
which is $\Delta n=1$ in these two cases. These behaviors are obviously independent of disorder averaging in noninteracting systems.

\subsection{Disorder and energy dependence of the occupation spectrum}

\begin{figure}[t]
\includegraphics[width=\columnwidth]{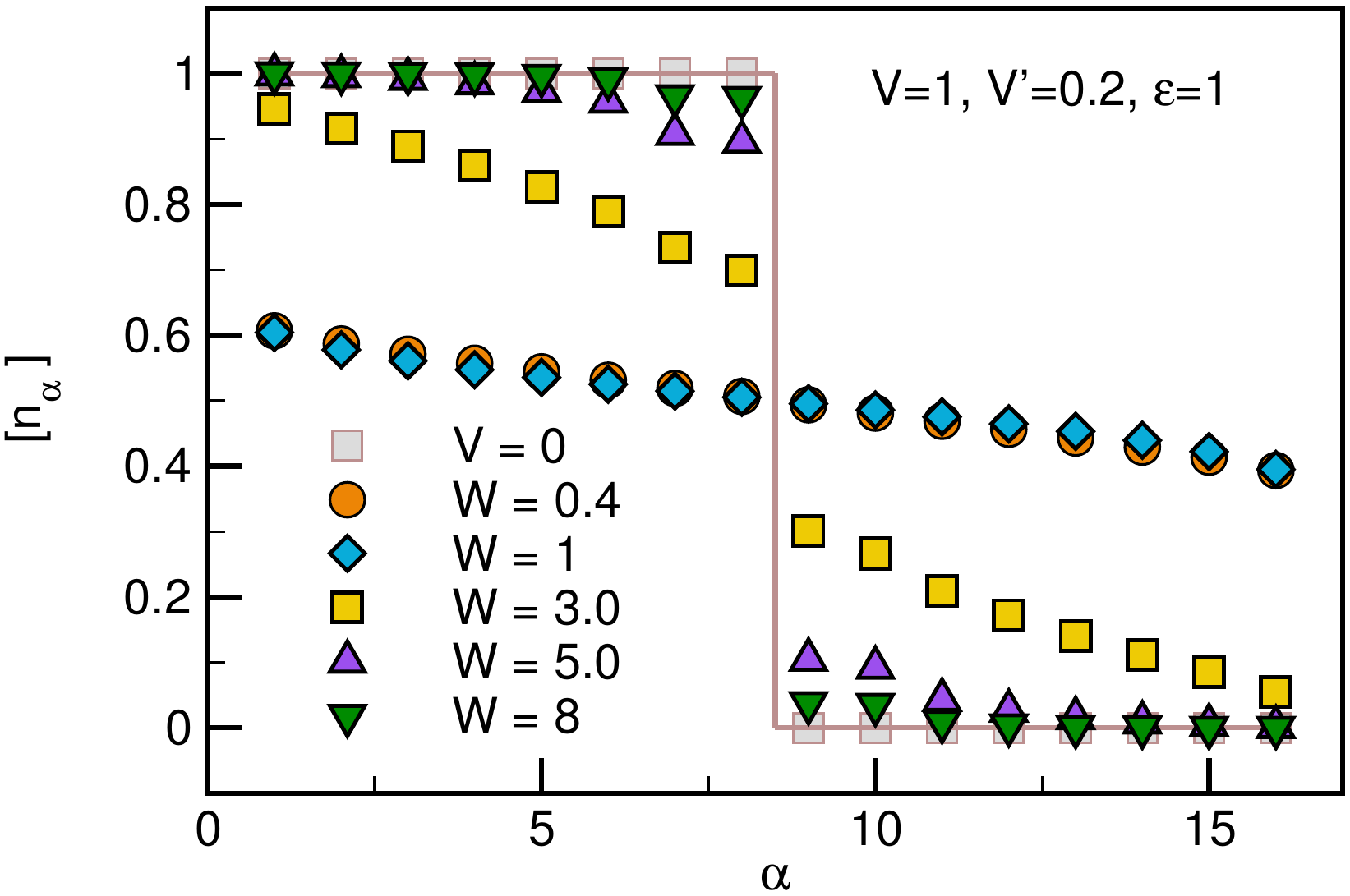}
\caption{(Color online)
Disorder-averaged occupation spectrum for different disorder strength. $L=16$, $V=1$, $V'=0.2$, $\varepsilon=1$.}
\label{fig:nalpha}
\end{figure}

\begin{figure}[t]
\includegraphics[width=\columnwidth]{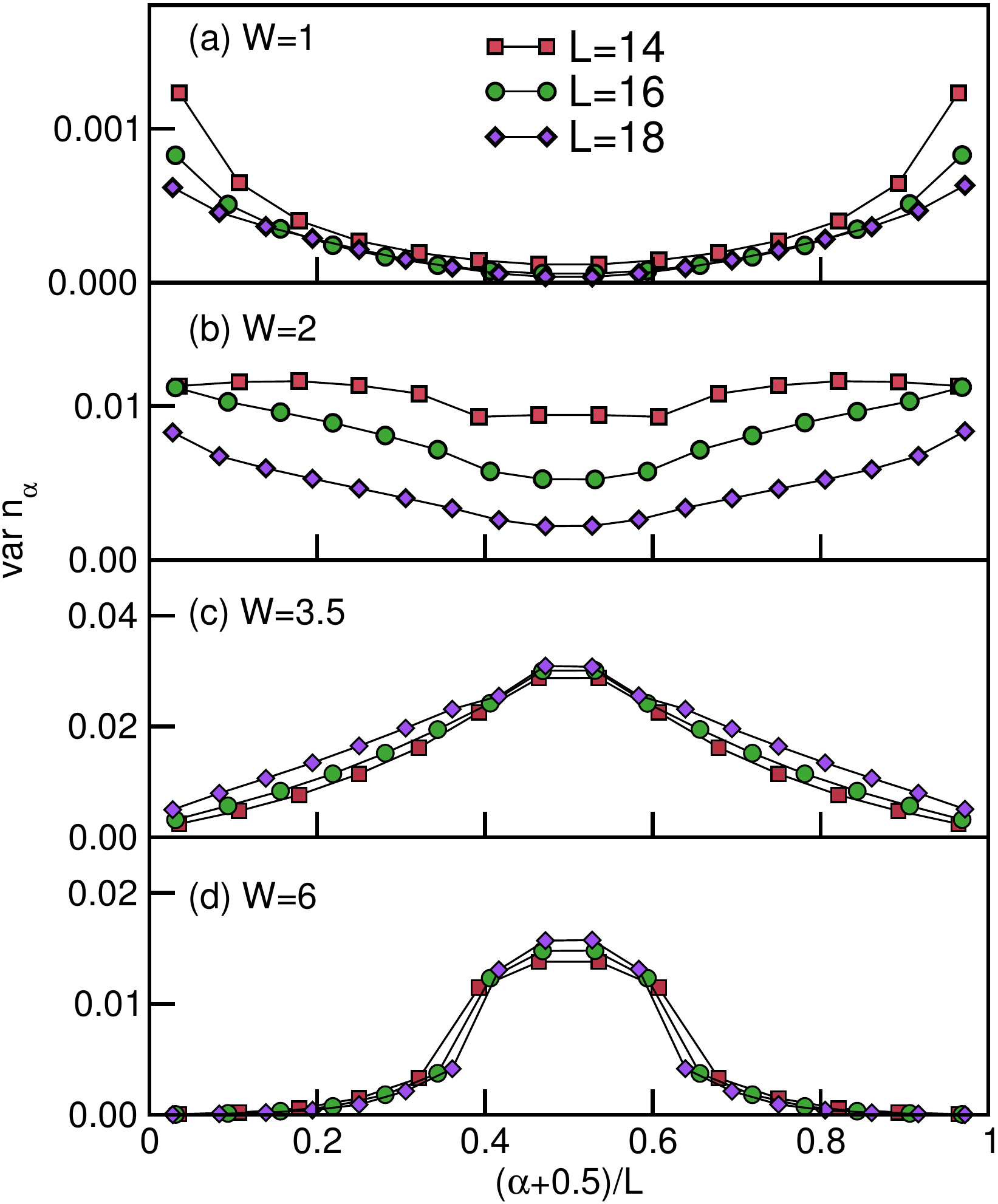}
\caption{(Color online)
Variance of the OPDM eigenvalues versus $(\alpha+0.5)/L$ for (a) $W=1$, (b) $W=2$, (c) $W=3.5$, (d) $W=6$ for $L=14,16,18$ ($V=1$, $V'=0$).}
\label{fig:var_nalpha}
\end{figure}

\subsubsection{Disorder-averaged occupation spectrum and its $W$-dependence}

We now turn to a discussion of the overall $W$-dependence of the occupation spectrum, considering 
disorder averaged quantities first (disorder averages are indicated by $\lbrack \cdot  \rbrack$). 
At finite values of $t,V$ in the MBL phase, the 
discontinuity $\lbrack \Delta n \rbrack $ remains finite but smaller than one, while some $\lbrack n_\alpha\rbrack $ are correspondingly smaller as well.
Far away from the ``Fermi-edge" at $\alpha_N$, $\lbrack n_\alpha\rbrack  \approx 1$ or  $\lbrack n_\alpha\rbrack  \approx 	0 $.
This implies that the corresponding quasiparticles  have a very simple structure, with very little contributions beyond the single-particle space.
As Fig.~\ref{fig:nalpha} shows, this behavior persists in the entire MBL phase $W \gtrsim 3.5$.
It also penetrates into the ergodic phase, at least on small finite systems (see the discussion below).

Sufficiently deep in the ergodic phase (see the data shown in Fig.~\ref{fig:nalpha} for $W=0.4$ and $W = 1$), $\lbrack n_\alpha\rbrack $ becomes
a smooth function of $\alpha$ and the discontinuity $\Delta n$ is small and just as large as other differences
$\lbrack n_{\alpha+1} - n_\alpha \rbrack$.

\subsubsection{Realization-to-realization fluctuations}
 
It is quite instructive to study both the results for $n_\alpha$ on individual realizations and the entire
distributions $\mathcal{P}(n_\alpha)$ obtained from averaging over many realizations.
We showed in Ref.~\onlinecite{Bera2015} that $\mathcal{P}(n_\alpha)$ develops a characteristic structure in the MBL phase with two maxima at $n_\alpha =0 $ and $n_\alpha =1$, consistent with the preceding discussion.
In Fig.~\ref{fig:var_nalpha}, we plot the variance of the OPDM eigenvalues versus $(\alpha+1/2)/L$. In the ergodic regime, the overall fluctuations
are small and are the largest at the edges. At the transition (see the data shown for $W=3.5$ in Fig.~\ref{fig:var_nalpha}(c)), the situation changes as the 
largest fluctuations now occur at the emergent Fermi edge (i.e., $(\alpha+1/2)/L=0.5$). This behavior persists in the MBL phase, where the overall fluctuations
decrease again, while their maximum remains at $(\alpha+1/2)/L=0.5$ (see the data shown for $W=6$ in Fig.~\ref{fig:var_nalpha}(d)).

\begin{figure}[t]
\includegraphics[width=\columnwidth]{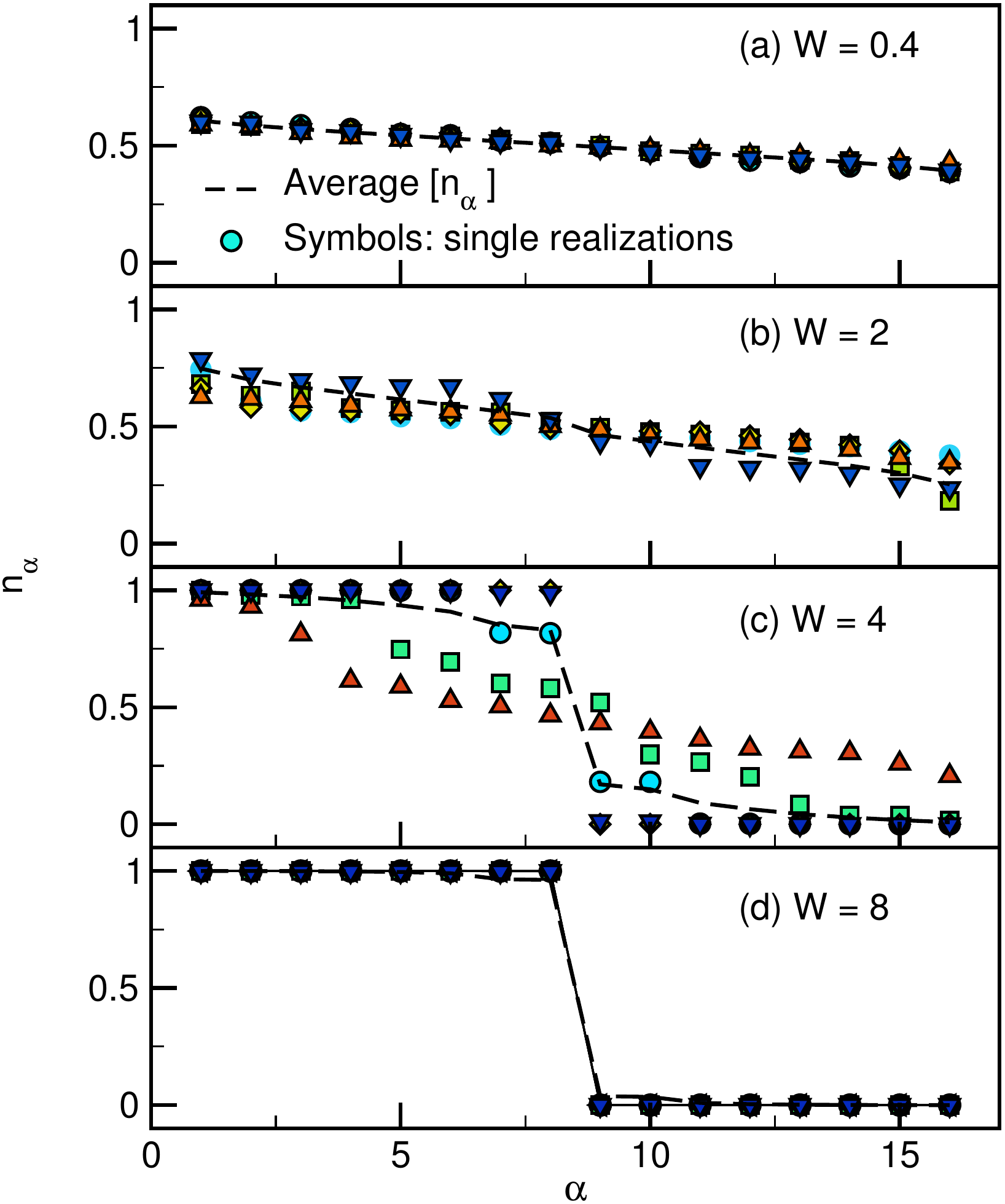}
\caption{(Color online)
Occupation spectrum in individual realizations for different disorder strength. $L=16$, $V=1$, $V'=0.2$, $\varepsilon=1$: (a) $W=0.4$, (b) $W=2$, (c) $W=4$, (d) $W=8$.}
\label{fig:real}
\end{figure}

By inspecting individual realizations as well as such distributions, we find that deep in the MBL phase,
virtually all realizations look quantitatively and qualitatively similar (i.e., they exhibit a discontinuity
$\Delta n$ with small fluctuations), as is shown in Fig.~\ref{fig:real}(d). 
Upon lowering $W$ and by approaching the transition (see Fig.~\ref{fig:real}(c) for an example), for some realizations, the distributions appear to be smooth, with no discernible discontinuity, while for the majority of cases, the typical MBL behavior prevails. 
We can understand this observation from the fact that in the vicinity of the transition, increasingly large thermal clusters are expected to exist in an overall insulating system
\cite{Gopalakrishnan2015,Vosk2015}.
Conversely, on the ergodic side, occasionally we still find realizations with a discontinuity, while the majority of realizations now has smooth occupation spectra (see Fig.~\ref{fig:real}(b) for an example).
Sufficiently far away from the transition and in the ergodic phase, practically all realizations result in smooth spectra (see Fig.~\ref{fig:real}(a) for an example).

\subsubsection{State-to-state variations in individual realizations}

\begin{figure}[t]
\includegraphics[width=\columnwidth]{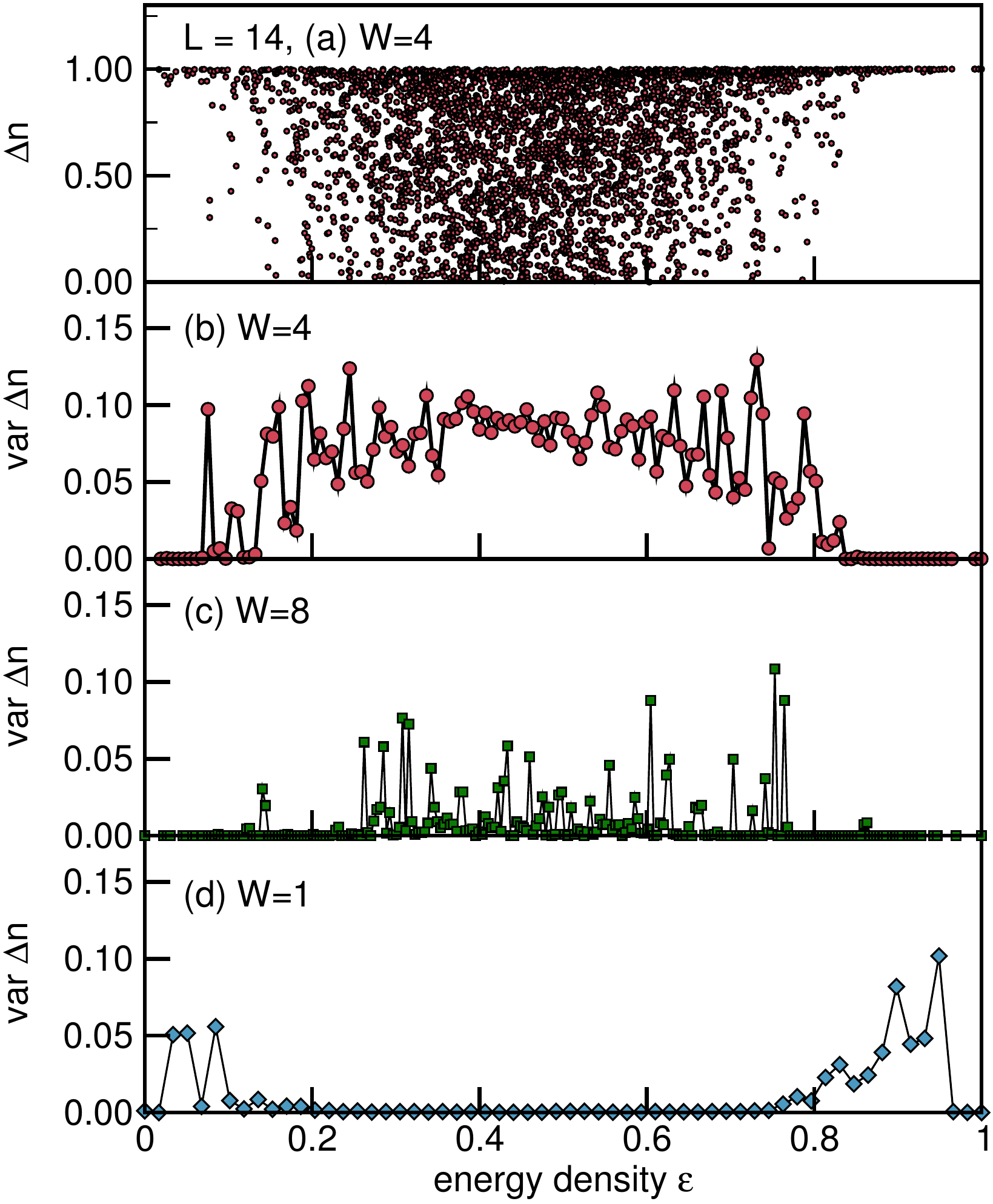}
\caption{(Color online)
State-to-state variation of the discontinuity $\Delta n$ for a
single realization at $W=4$, $V=1$, $V'=0.2$: (a) $\Delta n$ versus energy density $\epsilon$, (b) var $\Delta n$ for $W=4$, (c) var $\Delta n$ for $W=8$, (d) var $\Delta n$ for $W=1$,
(in (b)-(d), we use a binning of $\Delta E=0.2 $ on the energy axis).}
\label{fig:statetostate}
\end{figure}

It is important to stress that there are not only variations in the occupation spectrum from realization to realization (that are typically large in the transition region), but also from eigenstate to eigenstate for a given realization.
In order to illustrate this, we plot the discontinuity $\Delta n$ versus energy density in Fig.~\ref{fig:statetostate}(a), computed in all eigenstates for $W=4$,
which is close to the transition and on the MBL side.
At the edges of the spectrum, the discontinuity is $\Delta n =1$, with virtually no fluctuations, implying that the many-body states at low energies have a particularly simple structure in Fock space. 
In the bulk of the spectrum, the fluctuations of $\Delta n$ become large and exhaust the full range $0 \lesssim \Delta n \leq 1$.

Figures~\ref{fig:statetostate}(b)-(d) show the variance of the discontinuity for a fixed realization as a function of energy (we use a binning of $\Delta E=0.2$) 
for $W=4,8,1$, respectively.
The data for $W=4$ in (b) are obtained from the data shown in (a), and reflect the observations made before in a more quantitative way: vanishing fluctuations at the edges that become
large in the bulk of the spectrum. Deep in the MBL phase, the fluctuations tend to decrease (see Fig.~\ref{fig:statetostate}(c)). 
The fact that, in the MBL phase, state-to-state fluctuations of the discontinuity are the largest in the bulk but vanish at the edges of the spectrum can 
be viewed as a manifestation of the notion that many-body effects in the presence of disorder become the most relevant in the bulk of the system, i.e., at high energy densities.
This conforms with the idea that MBL is primarily a  high-energy phenomenon that 
 happens in what usually is considered the nonuniversal part of the many-body spectrum (that fact notwithstanding, the low-energy portions still belong to the MBL phase).
In the ergodic phase, however, the behavior is quite different and conforms with the expectation for a generic system that obeys ETH: in the bulk, the 
state-to-state fluctuations are small and are the largest at the edges (see Fig.~\ref{fig:statetostate}(d)).

To further analyze those states with a small $\Delta n$, we search for a correlation with eigenstates with a large entanglement. 
More generally speaking, the question is whether the tails in the distribution of the entanglement entropy \cite{Vosk2015,Luitz2016,Lim2016}
that appear as the transition is approached can be related to the tails in the distribution of $\Delta n$.
The entanglement entropy is defined as 
\begin{equation}
S_{\rm vN} = -\mbox{tr}_R \left (\rho_R \ln\rho_R \right),
\end{equation}
where $\rho_R$ is the reduced density matrix of the right half of the system with $L/2$ sites.

\begin{figure}[t]
\includegraphics[width=\columnwidth]{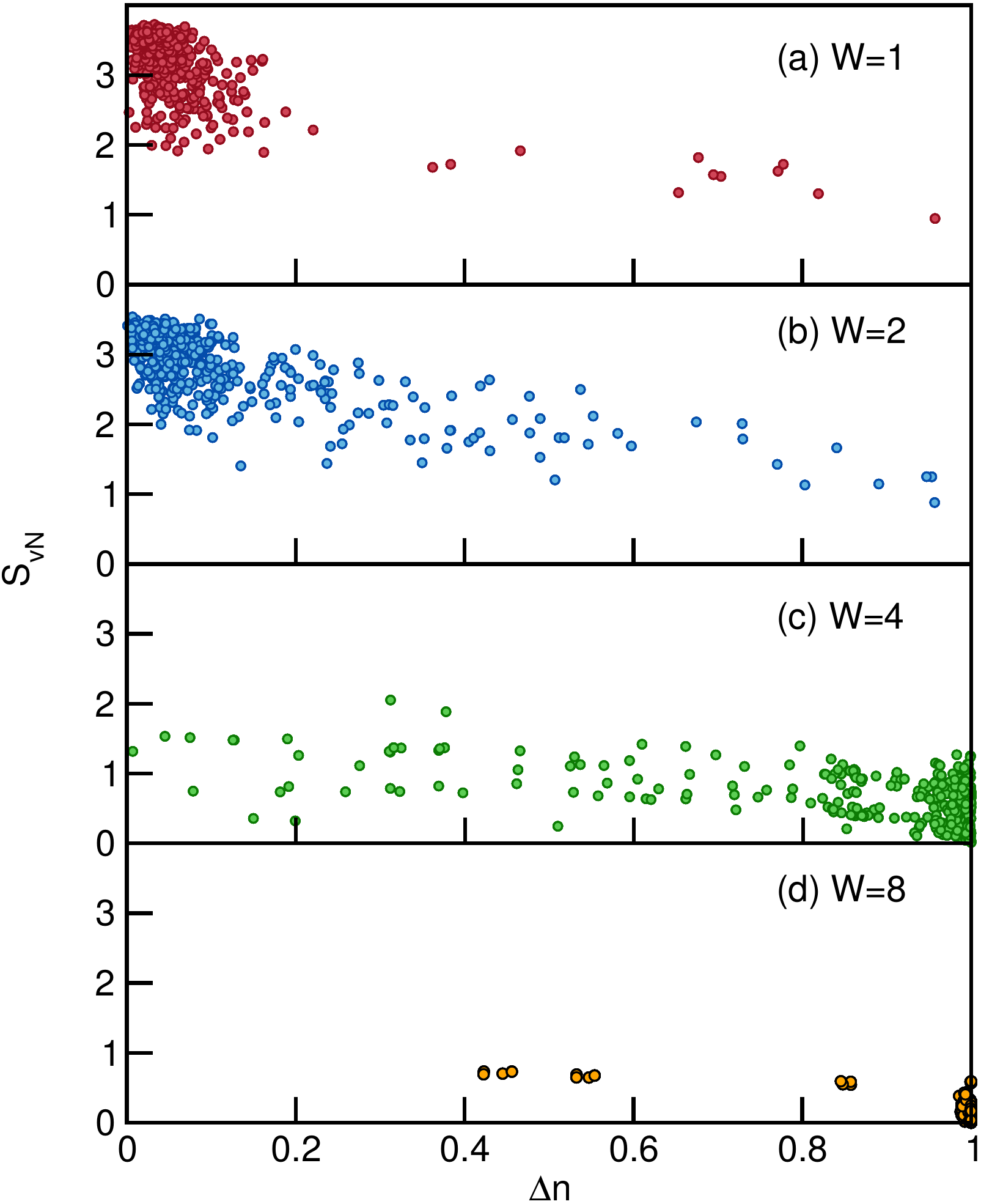}
\caption{(Color online)
Half-cut entanglement entropy $S_{\rm vN}$ versus occupation-spectrum discontinuity $\Delta n$ computed for a given disorder realization
in all many-body eigenstates for $L=14$ and (a) $W=1$, (b) $W=2$, (c) $W=4$, and (d) $W=8$ ($V=1$, $V'=0$).
\label{fig:svn_delta}}
\end{figure}

To clarify this, again for a fixed realization, we plot the half-cut entanglement entropy $S_{\rm vN}$ versus the occupation-spectrum discontinuity $\Delta n$
in all many-body eigenstates for $L=14$ and different disorder strengths $W=1,2,4,8$ in Figs.~\ref{fig:svn_delta}(a)-(d), respectively.
The results confirm the overall expectation that large-entanglement eigenstates have small discontinuities $\Delta n$
and vice versa. 
This trend is most clearly visible in the data for the ergodic phase shown in Figs.~\ref{fig:svn_delta}(a) and (b).
In the MBL phase (see the data for $W=4,8$), there are less cases with large $S_{\rm vN}$ overall 
(keeping in mind that the plots show results for only 3432 states in one realization).
The trend is still present in the data that states with a small $\Delta n$ have larger values of $S_{\rm vN}$, while the most striking
difference between ergodic (see the data for $W=1$) and MBL phase (see the data for $W=8$) is the bunching of points at  small $\Delta n$ and
large $S_{\rm vN}$ versus large $\Delta n$ and small $S_{\rm vN}$.

\subsubsection{Distributions as a function of disorder strength}
 
\begin{figure}[tb!]
\includegraphics[width=\columnwidth]{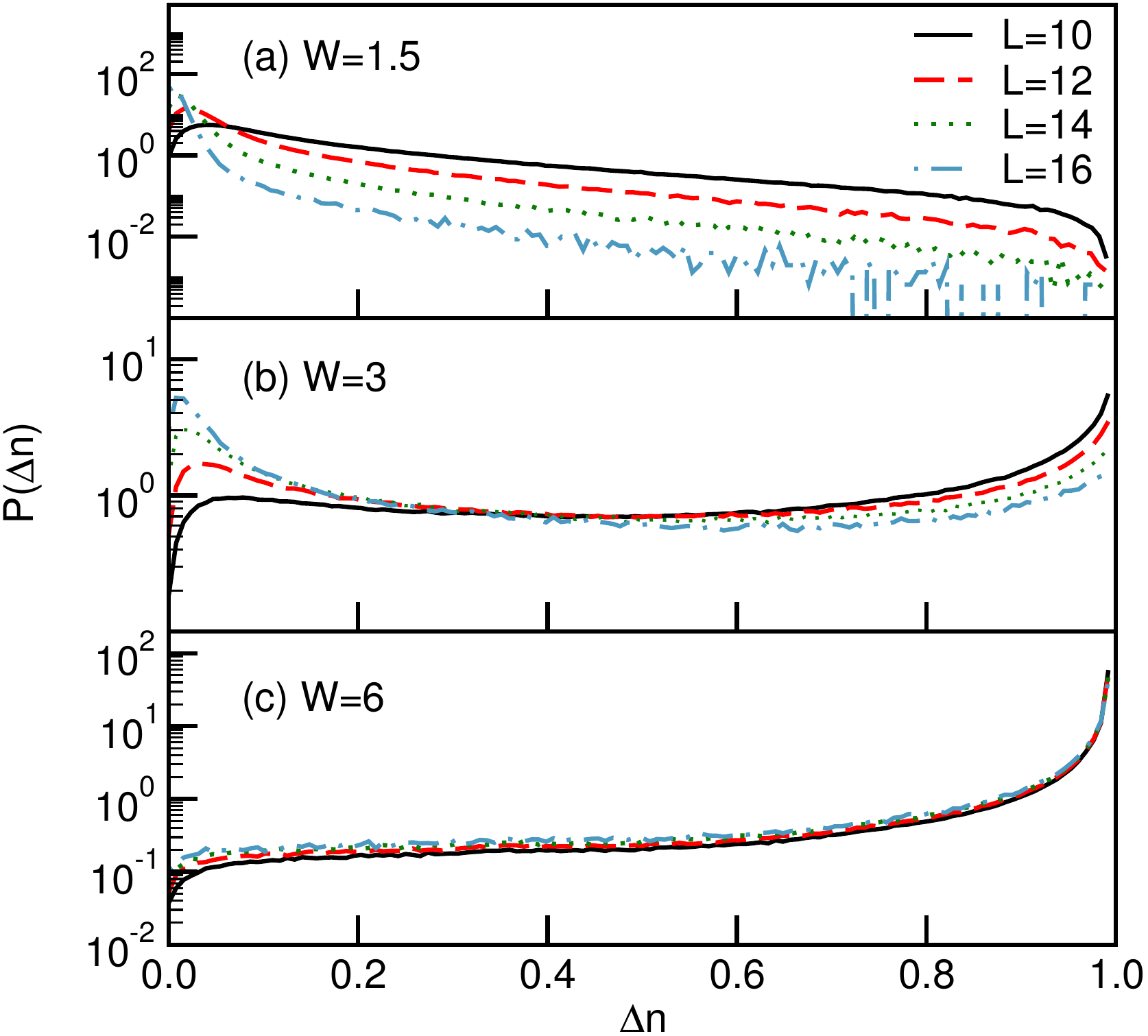}
\caption{(Color online) Distribution of the occupation spectrum discontinuity $\Delta n$ for different values of $W$ and $L$ for $V=1$, $V'=0$. 
}
\label{fig:pnalpha}
\end{figure}

One can further analyze the full distribution $\mathcal{P}(\Delta n)$ of the discontinuity as a function of disorder strength.
Figure~\ref{fig:pnalpha} shows the evolution of this distribution as one goes from the ergodic phase ($W=1.5$) through the transition region ($W=3$)
into the MBL phase ($W=6$). Clearly, in the ergodic phase, the maximum is at $\Delta n \approx 0$ (plus $1/L$ corrections). In the transition region,
$\mathcal{P}(\Delta n)$ is much broader and there are two maxima at $\Delta n \approx 0$ and $1$. As system size increases, the maximum at $\Delta n=0$ becomes more pronounced
while the one at $\Delta n=1$ decreases, consistent with the prediction \cite{Luitz2015} that $W=3$ is on the ergodic side of the transition. 
Note that such a bimodal structure in the distributions has also been observed for local observables \cite{Luitz2016}.
In the MBL phase, the 
distribution is sharply peaked at $\Delta n=1$, as expected.

\begin{figure}[tb!]
\includegraphics[width=1\columnwidth]{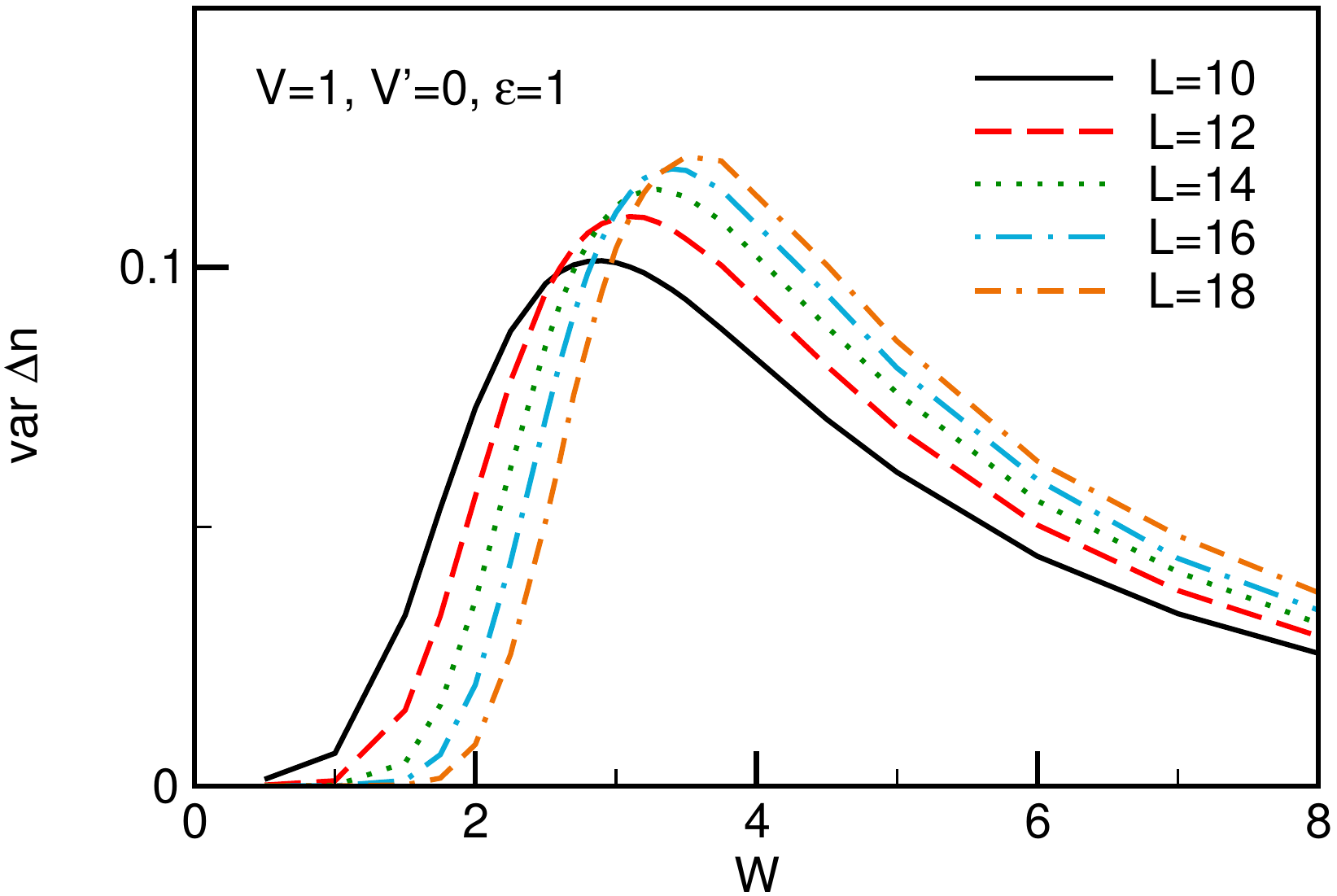}
\caption{(Color online) Variance of the  distribution of the discontinuity as a function of $W$ for different values of $L$ at $V=1$, $V'=0$ and $\epsilon=1$.}
\label{fig:varDeltaN}
\end{figure}

Additional information can be obtained from plotting the variance var($\Delta n$) of $\mathcal{P}(\Delta n)$ versus $W$, which is shown in Fig.~\ref{fig:varDeltaN}.
These data unveil an interesting finite-size dependence: for small $W$, var($\Delta n$) decreases with $L$, while in the MBL, it increases.
Moreover, there is a maximum of  var($\Delta n$)  in the transition region. 
Interpreting the finite size dependence is complicated by the fact that var$(n_\alpha)$ depends on $\alpha$ and has a peak at $\alpha/N=1/2$; therefore, as $L$ increases and as one obtains a finer resolution of the $\alpha$ axis, one gets closer to the peak and therefore the fluctuations can increase (see the data for  var $n_{\alpha}$ versus $\alpha$ shown in Fig.~{\ref{fig:var_nalpha}}).
This variance contains similar information as the occupation entropy that will be discussed in Sec.~\ref{sec:occ_entropy}.

\subsection{Occupation spectrum in the ergodic phase and thermalization}
\label{sec:nthermal}

In the ergodic phase, we expect the occupation spectrum  $\lbrack n_{\alpha} \rbrack $ to be smooth, with the actual values determined by the temperature.
Since in the ergodic phase, the eigenstate thermalization hypothesis should hold, the set of $n_{\alpha}$ computed from individual many-body eigenstates should be identical to the expectation values in the canonical ensemble.
There is a minor caveat here since the operators  $\hat n_\alpha = c_\alpha^\dagger c_{\alpha}$ are not necessarily local in the ergodic phase, in contrast to the MBL phase, where $\hat n_\alpha$ can be related to the quasi-local conserved quantities (see the discussion in Sec.~\ref{sec:lbits}).
Thus, in the MBL phase the index $\alpha$ can be thought of as a position index, as argued above, while in the ergodic phase, this is not the case since the natural orbitals are extended wave functions.
Despite the nonlocality of the $\hat n_\alpha$, their eigenstate expectation values are still expected to be thermal in the ETH sense, in analogy to the behavior of the quasimomentum distribution $n_k$ of a clean system  \cite{Rigol2008,Wright2014,Sorg2014}.

In order to verify this expectation, we first consider individual realizations $\rho$ (corresponding to a fixed set of $L$ local potentials $\epsilon_i$) and compute a temperature $T_{\rho}$  from the condition
\begin{equation}
E_{\rho} = \mbox{Tr} \lbrack \rho^{(1)}_{\rm can}(T_\rho) H_{\rho} \rbrack\,.
\end{equation}
Here, $E_{\rho}$ is the energy of the many-body eigenstate closest to the target energy $E$ from which we extracted the natural orbitals and the occupation spectrum;
$\rho_{\rm can}(T_{\rho}) = e^{-H_\rho / T_\rho} /Z_\rho$ is the canonical ensemble with the partition function $Z_\rho = \mbox{Tr}( e^{-H_\rho / T_\rho})$.
We then compute the thermal expectation value of the OPDM at that temperature from 
\begin{equation}
\rho^{(1)}(T_\rho) = \mbox{Tr}\lbrack  \rho_{\rm can}(T_\rho) c_i^\dagger c_j   \rbrack \, 
\end{equation}
and diagonalize it to obtain the thermal expectation values of the occupation spectrum $n_{\alpha}(T_\rho)$.
Next, we average over many disorder configurations, resulting in $ \lbrack n_{\alpha}(T) \rbrack$ (where we have now dropped the index $\rho$).
Clearly, this procedure results in a distribution of temperatures $T_\nu$. 
We observe that these distributions of temperatures are quite narrow at weak disorder and in the ergodic phase, while upon approaching the transition and in the MBL phase, they become increasingly broad and develop tails \cite{Martynec2015}.

\begin{figure}[tb]
\includegraphics[width=\columnwidth]{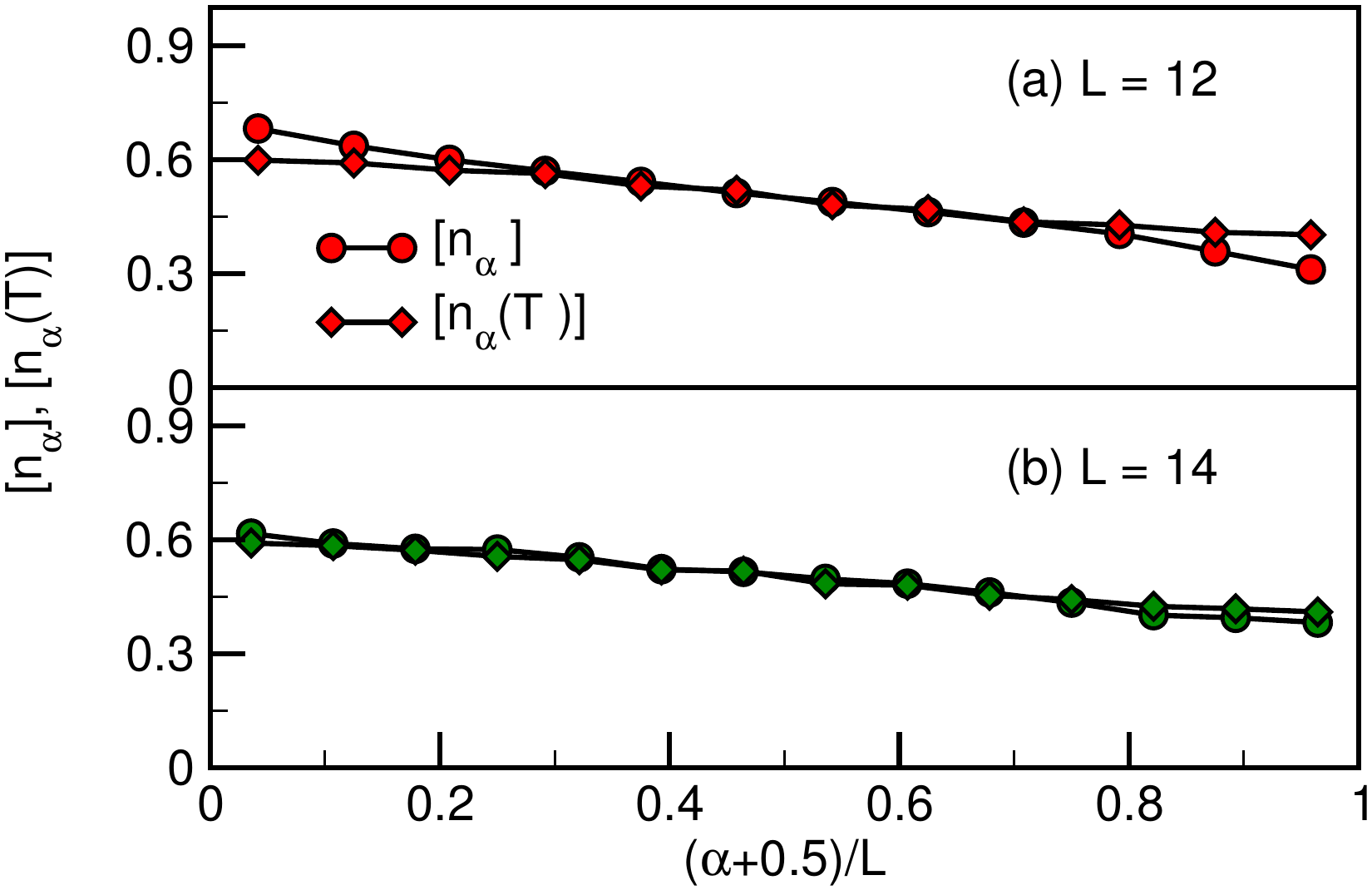}
\caption{(Color online)
Comparison of $[n_\alpha]$ to $[n_\alpha(T)]$. (a) $L=12$, (b) $L=14$. $W=0.4$, $V=1$, $V'=0.2$, $\varepsilon=1$.
}\label{fig:nthermal}
\end{figure}

A direct comparison of the disorder-averaged occupation spectrum computed from either individual many-body eigenstates
or in the canonical ensemble is shown in Figs.~\ref{fig:nthermal}(a) and (b) for $L=12$ and $14$, respectively. The agreement is very good, while the quantitative
differences between $\lbrack n_\alpha \rbrack$ and $\lbrack n_\alpha(T) \rbrack$
 are of the magnitude typical for such a small system \cite{Sorg2014}. 
We studied these differences as a function of system size (not shown here) and find that they rapidly decrease with $L$, as suggested by the comparison of the $L=12$ and $L=14$
data.
As a consequence, the difference $\Delta n$ between the occupation of the $N$th and $N+1$st natural orbital should 
vanish with system size and must be strictly zero in the ergodic phase in the thermodynamic limit.
To summarize, computing $\Delta n$ serves as a very clean measure of obtaining the 
phase diagram. 

\subsection{Phase diagrams from the occupation spectrum discontinuity}

Our previous discussion suggests that the discontinuity $\Delta n$ can be used to
compute phase diagrams of interacting disordered systems. We expect $\Delta n  \to 0 $ in the ergodic phase
as $L$ increases and $\Delta n >0 $ in the MBL phase.

\begin{figure}[tb!]
\includegraphics[width=\columnwidth]{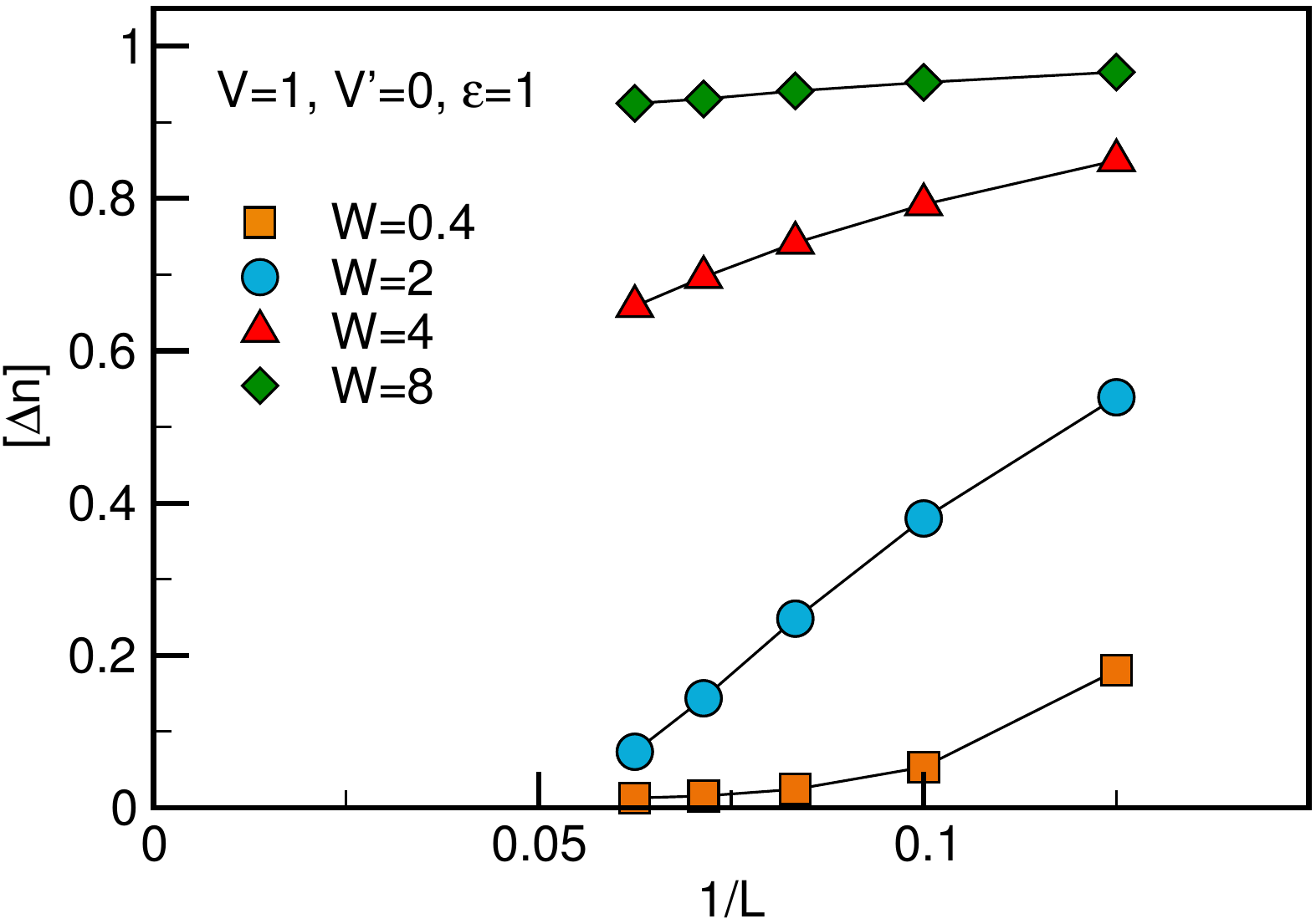}
\caption{(Color online)
Finite-size dependence of the disorder-averaged occupation-spectrum discontinuity for $W=0.4, 2,4,8$ and $V=1$, $V'=0$, $\varepsilon=1$.
}
\label{fig:Ldep}
\end{figure}

Currently, we are limited to the system sizes accessible to exact diagonalization. We demonstrate that the 
finite-size dependence of $\Delta n$ deep in either phase is consistent with our expectations described above.
Figure~\ref{fig:Ldep}, where we plot $\Delta n$ versus $1/L$, shows that this is indeed the case for $W=1$ and $W=8$.
For intermediate values of $W$ closer to where one expects the transition, the trend is not clear yet, 
due to the small system sizes.
Note, however, that a somewhat faster convergence of $\Delta n$ to zero is observed when  plotting the 
median instead of the mean of the distribution $P(\Delta n)$. While we have not done so here, where we are mainly
interested in qualitative properties, for a quantitative analysis, it may thus be better to study the median.


\begin{figure}[tb!]
\includegraphics[width=1\columnwidth]{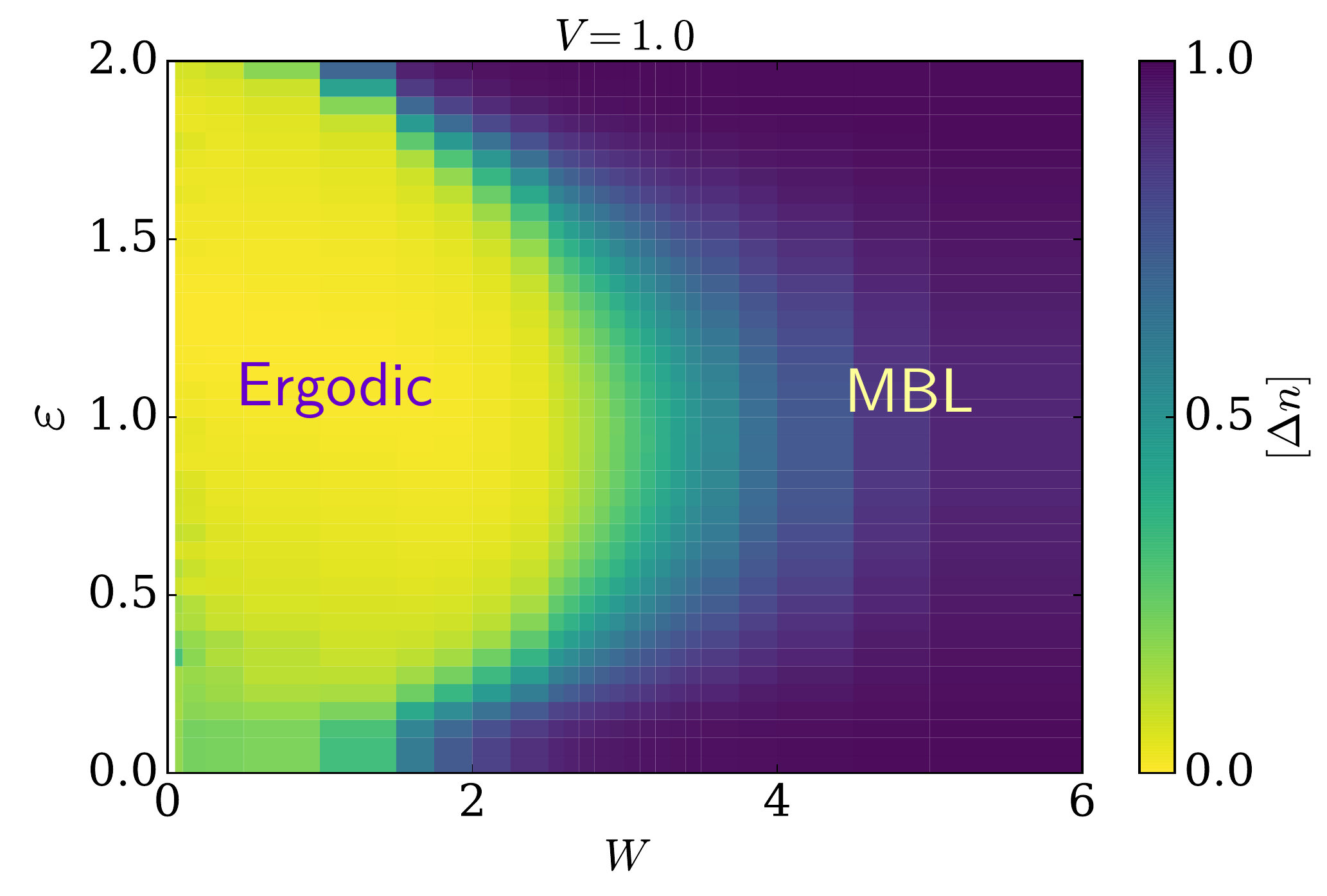}
\caption{(Color online)
Phase diagram, obtained from plotting the disorder average $\lbrack \Delta n\rbrack$ of the occupation-spectrum discontinuity,  of interacting 1D spinless fermions with uncorrelated diagonal disorder Eq.~\eqref{eq:ham} obtained from plotting $\Delta n = n_{N+1} - n_N$
as a function of  $\epsilon$ and $W$ at $V=1$, $V'=0$.
ED data for $L=16$ and $N=8$ particles.
\label{fig:phasediag}
}
\end{figure}

\begin{figure*}[tb!]
\includegraphics[width=1\textwidth]{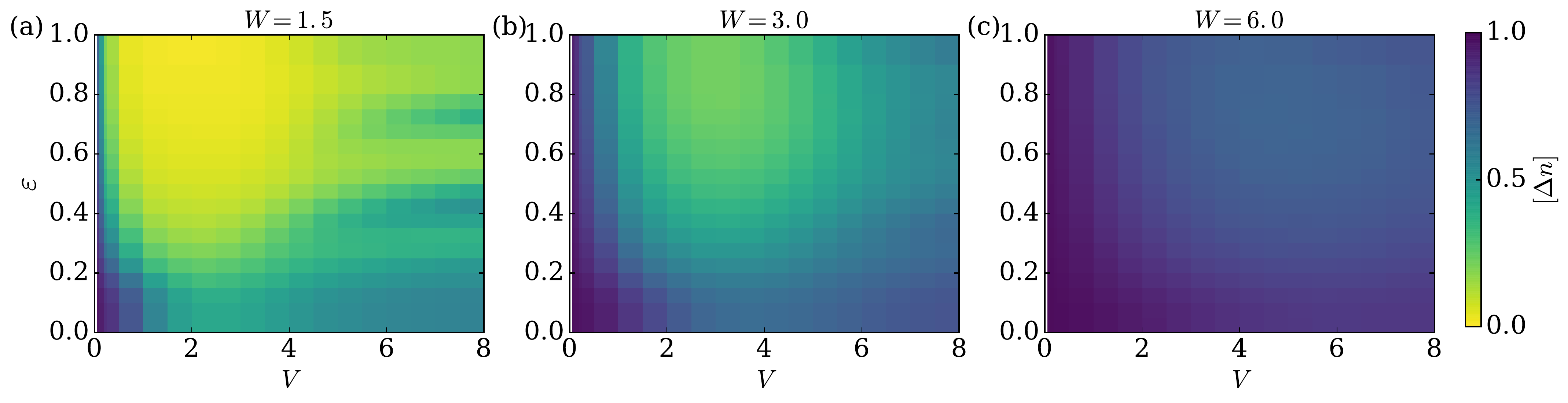}
\caption{(Color online) Phase diagram of interacting 1D spinless fermions with uncorrelated diagonal disorder Eq.~\eqref{eq:ham} in the $\varepsilon$-$V$ plane for $L=16$ for three different values of disorder: (a) $W=1.5$, (b) $W=3$, and (c) $W=6$ ($V'=0$); obtained from plotting the disorder average $\lbrack \Delta n\rbrack$ of the occupation-spectrum discontinuity.} 
\label{fig:phasediag_a}
\end{figure*}

Nonetheless, it is very instructive to plot $\lbrack \Delta n \rbrack$ as a function of energy density $\varepsilon$ and disorder strength $W$  at a fixed $V=t$ (Fig.~\ref{fig:phasediag}) 
for a system size of $L=16$.
A key observation is that already these finite-size data produce the known features of the phase diagram of spinless fermions with short range repulsive interactions: the mobility edge, the value for the critical disorder strength at  infinite temperature $3 \lesssim W_c \lesssim 4$, as well as the reentrant behavior as a function of interaction strength \cite{Luitz2015,Pal2010,BarLev2015}.
An important next step, which is beyond the scope of the present work, is to employ recently developed DMRG methods \cite{Khemani2016,Kennes2016,Lim2016,Pollmann2016,Yu:2015wya,Serbyn:2016gv} that are specifically tailored for the MBL phase, to compute the OPDM. 

Figure~\ref{fig:phasediag_a} shows the discontinuity in the energy density versus interaction strength plane, which is not usually shown.
For small $W$ (see Fig.~\ref{fig:phasediag_a}(a)), a large region is delocalized, in particular, as $\epsilon$ increases. 
For intermediate disorder strength $W=3$ (see Fig.~\ref{fig:phasediag_a}(b)), the delocalized region (indicated by small values of $\Delta n$ shrinks considerably, while at large disorder $W=6$ (see Fig.~\ref{fig:phasediag_a}(c)), the system is uniformly in the MBL phase.


\subsection{Single-particle occupation entropy}
\label{sec:occ_entropy}
From the occupation spectrum $n_{\alpha}$, one can define a single-particle occupation entropy
via
\begin{equation}
S_{\mathrm{occ}}' = -\frac{1}{N} \sum_{\alpha} n_\alpha \mbox{ln} (n_\alpha/N)\,.
\end{equation}

\begin{figure}[t]
\includegraphics[width=\columnwidth]{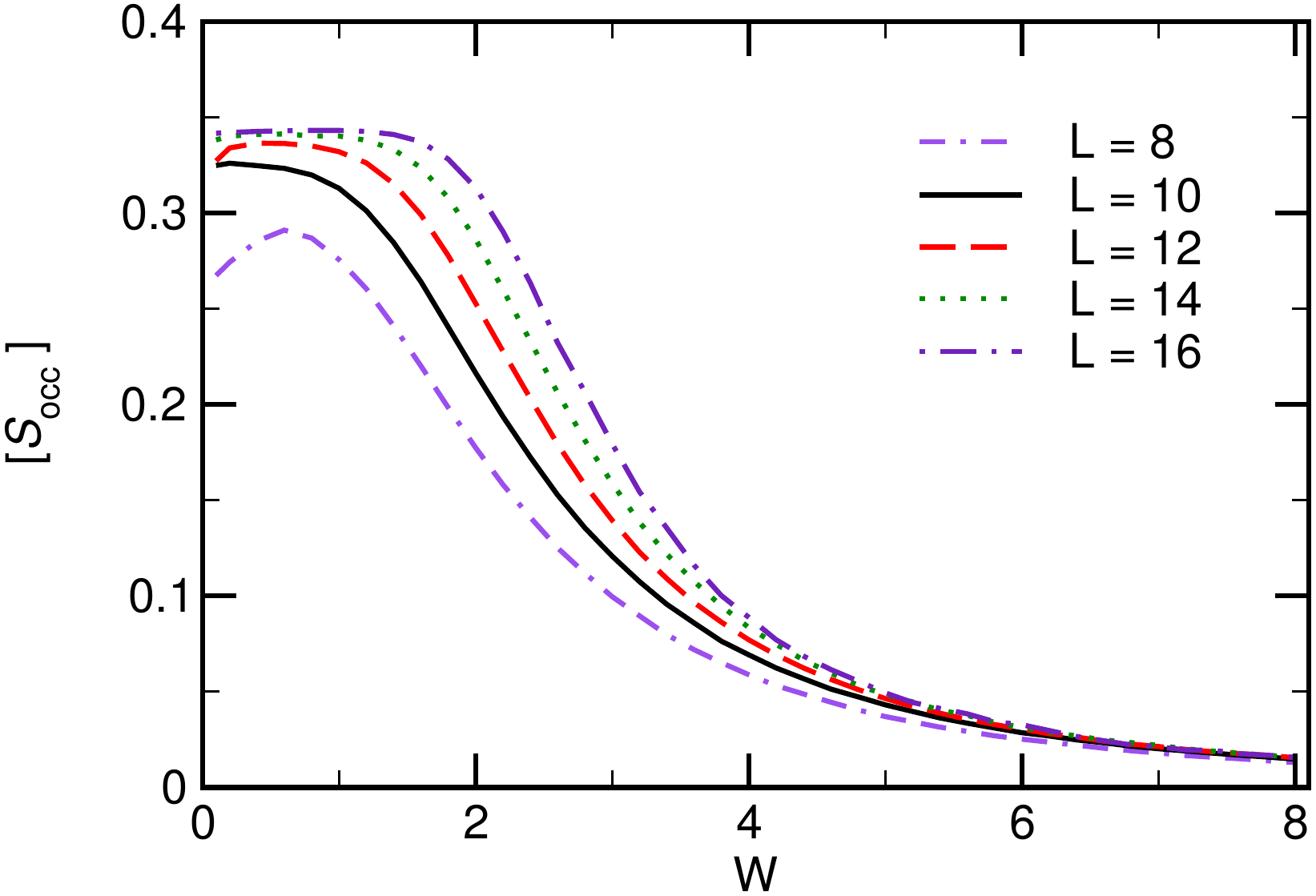}
\caption{(Color online)
Disorder-averaged single-particle occupation entropy $\lbrack S_{\rm occ} \rbrack$ for $V=1$, $V'=0.2$, $\varepsilon=1$ and for different system sizes  $L=8,10, \dots,16$.\label{fig:socc}}
\end{figure}

Note that here we choose a different normalization for the single-particle density matrix compared to Ref.~\cite{Bera2015}, namely
$\mbox{tr}[\rho^{(1)}]=1$ instead of $\mbox{tr}[\rho^{(1)}]=N$. As a consequence, the single-particle occupation
entropy is {\it intensive}. This normalization removes
a trivial overall volume-law dependence of $S_{\mathrm{occ}}'$.  Moreover, we also remove an additive 
term $ \mbox{ln}(N)$ which is the contribution of a Slater determinant such that $S_{\rm occ}=0$ for a product state:
\begin{equation}
S_{\mathrm{occ}}= S_{\mathrm{occ}}'-\mbox{ln}(N)= -\frac{1}{N} \sum_{\alpha} n_\alpha \mbox{ln} (n_\alpha)\,.
\end{equation}
The dependence of $S_{\mathrm{occ}}$ on disorder strength 
is illustrated in Fig.~\ref{fig:socc}(a) for several system sizes $L=8,10,12,\dots, 16$ and $V=1, V'=0.2$.
At small $W$, $S_{\mathrm{occ}}$ quickly becomes $L$-independent. The small downturn as $W\to 0$ seen for the smallest
system sizes such as $L=8,10$ is a remnant of the proximity to the integrable model ($W=0$, $V'=0$) and is quickly suppressed by increasing $L$
or making $V'$ larger.

For a fixed $L$ (apart from the small feature at small $W$), $S_{\mathrm{occ}}$ monotonically decreases with $W$ and becomes small in the MBL phase,
where it also exhibits a weak $L$-dependence only, saturating fast already for $L=16$.
The associated variance of the $S_{\mathrm{occ}}$ (fluctuations between different realizations), 
has a maximum in the transition region (data not shown here, see Ref.~\cite{Bera2015}). The position of this maximum is $L$-dependent and shifts towards $W_c$ as $W$ increases.

\subsection{Correlations between OPDM occupations computed in different many-body eigenstates}
\begin{figure*}[tbh]
\includegraphics[width=1\textwidth]{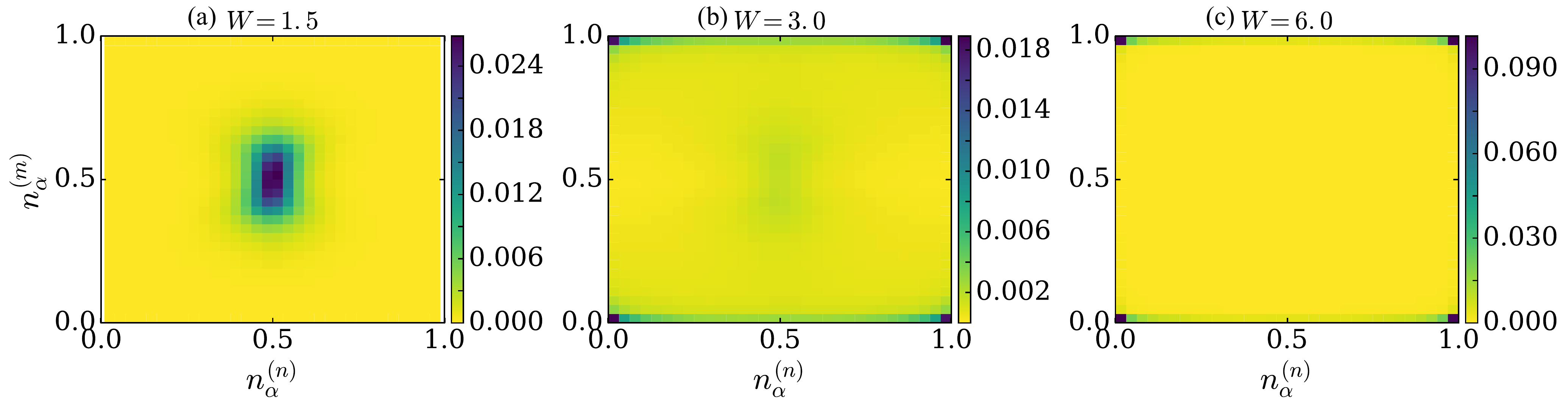}
\caption{(Color online) Probability density of obtaining a value $n_\alpha^{(m)}$ in state $|m\rangle$, given that a value $n_\alpha^{(n)}$ was obtained in state $|n\rangle$, with both being calculated with the operator $\hat{n}_\alpha$ obtained in the state $|n\rangle$. This plot is obtained for $L=14$, $V=1$, $V'=0$, taking 64 disorder realizations, 64 states $|n\rangle$ from the middle of the spectrum for each realization and then for each of these about 1500 states $|m\rangle$ distributed over the full energy spectrum.
}
\label{fig:nbeta}
\end{figure*}
When the Hamiltonian is given in terms of $L$ commuting conserved densities ${\hat n}^{(\rm qp)}_i$, as in Eq.~\eqref{eq:hlbit}, all eigenstates are product states of the form
\begin{equation}
	|n\rangle = \prod_{i=1}^L (c^{({\rm qp})\dagger}_i)^{{n}^{\rm (qp)}_i}|0\rangle,
	\label{eq:l-bitEigenstates}
\end{equation}
where the quasiparticle occupations ${n}^{\rm (qp)}_i=0,1$. 
Importantly, different eigenstates are product states of the same local operators. 
This suggests that the natural orbitals obtained from different eigenstates are close to each other, even though, as argued in Sec.~\ref{sec:lbits}, they are in principle state dependent.
To check this, we calculate the OPDM in eigenstate $|n\rangle$, diagonalize it with the unitary $U^{(n)}$, and construct the creation operator:
\begin{equation}
	c^\dagger_\alpha = \sum_{i}U^{(n)}_{\alpha,i}c^\dagger_i.
\end{equation}
These operators satisfy the relation $\langle n|c^\dagger_{\alpha}c_{\beta}|n\rangle = n_\alpha^{(n)}\delta_{\alpha,\beta}$ as before.
With this operator, we now take another eigenstate $|m\rangle$ and calculate the diagonal elements 
\begin{equation}
	n_\alpha^{(m)} = \langle m | c^\dagger_{\alpha}c_\alpha|m\rangle.
\end{equation}
If the natural orbitals are similar, then $n_\alpha^{(m)} \approx n_\alpha^{(n)}$ if $\alpha$ has the same occupation as in $|n\rangle$, but $n_\alpha^{(m)} \approx 1 - n_\alpha^{(n)}$ otherwise. 
We check this with the density plot of $n_\alpha^{(m)}$ versus $n_\alpha^{(n)}$ in Fig.~\ref{fig:nbeta}.
In the ergodic phase these occupations have a maximum at $0.5$ for both $|n\rangle$ and $|m\rangle$.
In the localized phase the occupations $n_\alpha^{(n)}$ are always close to either zero or one, as expected, and in the state $|m\rangle$ the occupations of $c_\alpha$ obtained from $|n\rangle$ indeed are predominantly close to either zero or one.
This demonstrates that the natural orbitals obtained for one eigenstate remain a good approximation of the natural orbitals computed in another state, and therefore the single-particle  content of the corresponding $l$-bit, in other eigenstates.
For $W \lesssim 3.5$, there is the additional effect that some states $|m\rangle$ are from below the mobility edge and thus the natural orbitals from states $|n\rangle$ in the middle of the spectrum (which are already ergodic) should not be correlated to those.

\section{OPDM properties in the interacting Aubry-Andr{\'e} model}
\label{sec:AA}
\begin{figure}[tb]
\includegraphics[width=1\columnwidth]{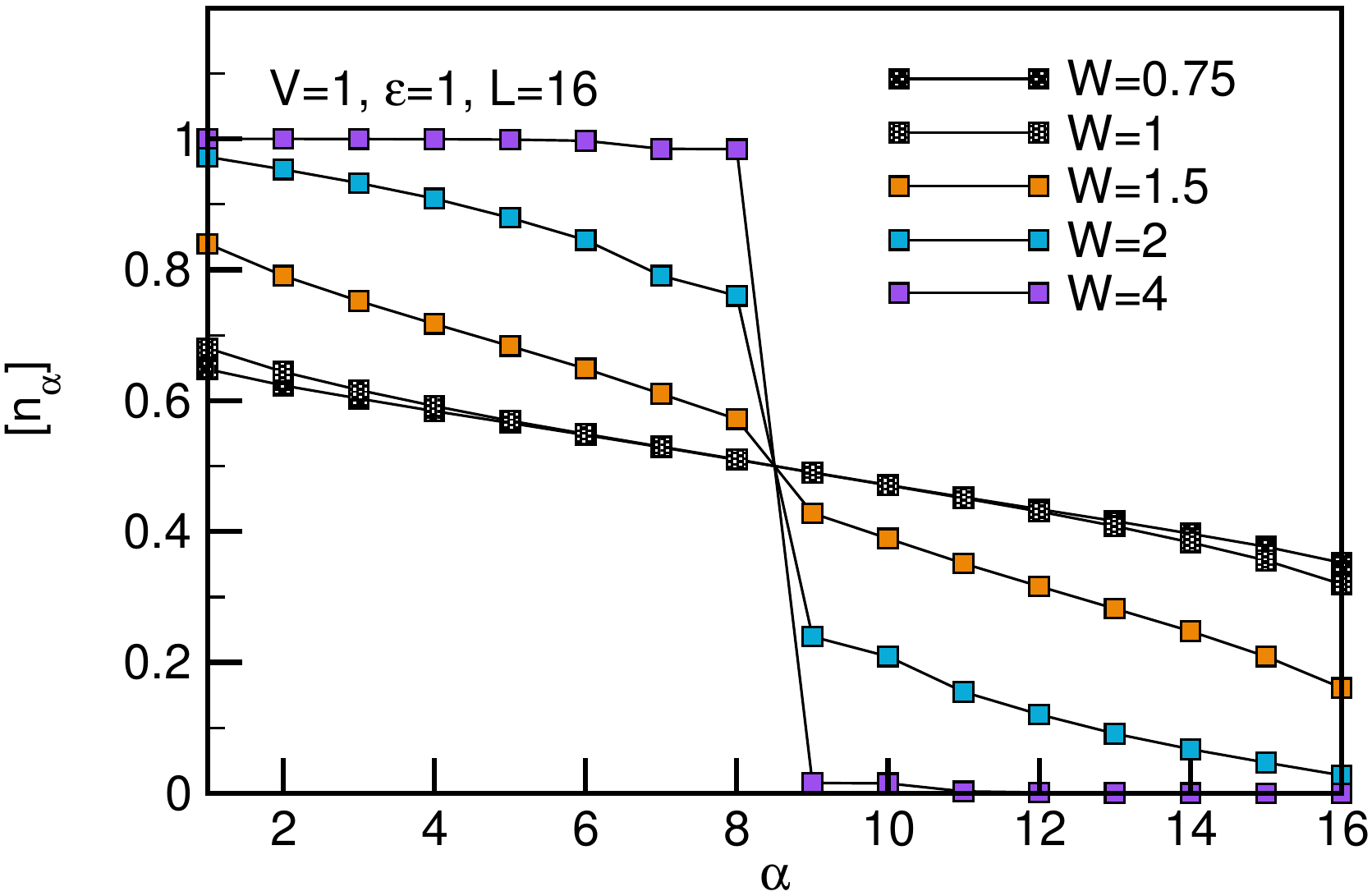}
\caption{(Color online) Disorder-averaged occupation spectrum in the interacting Aubry-Andr{\'e} model (spinless fermions) for different disorder strengths, system size $L=16$ and energy density $\varepsilon = 1$ ($V=1$, $V'=0$).}
\label{fig:occupationsAAH}
\end{figure}

\begin{figure}[tb]
\includegraphics[width=1\columnwidth]{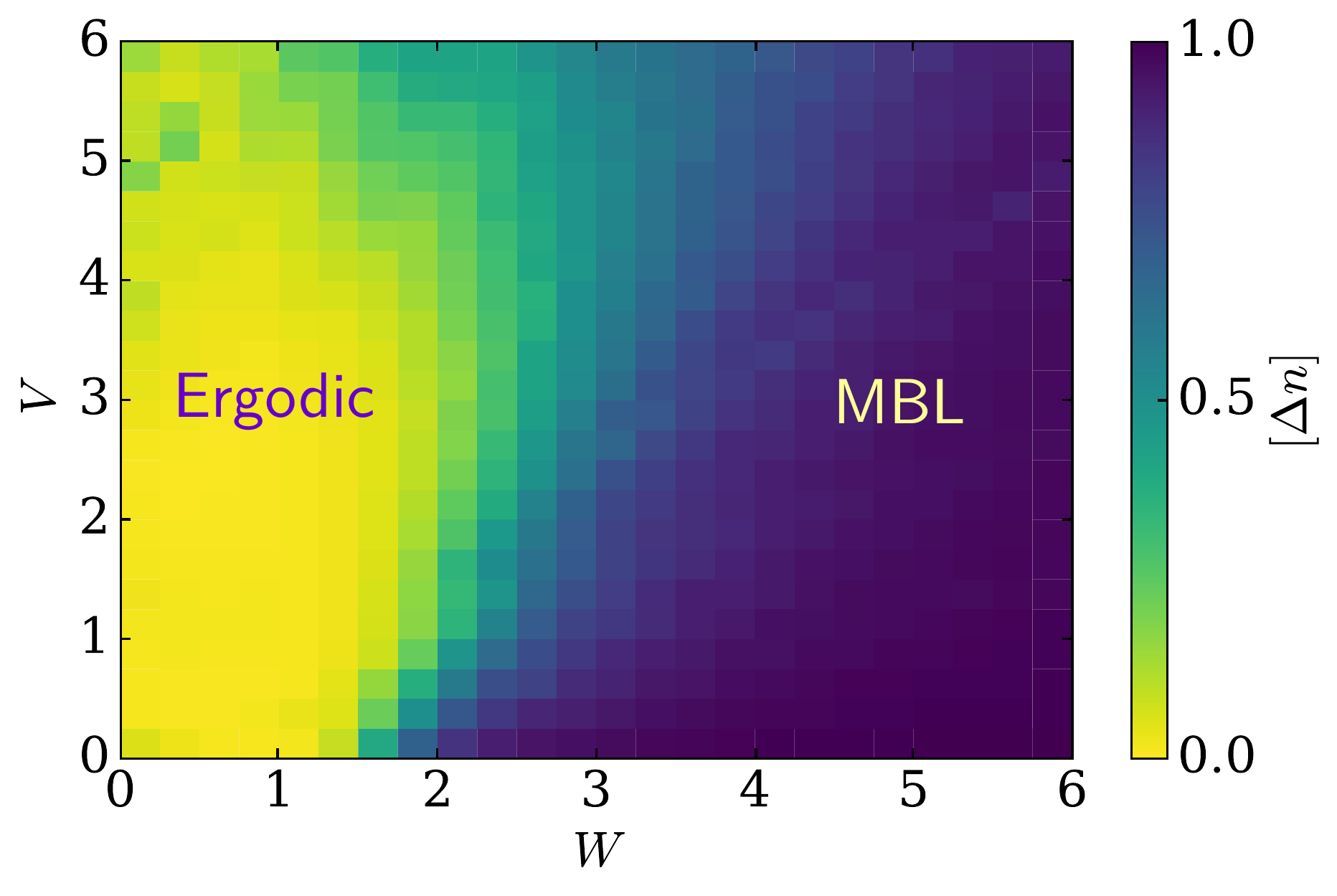}
\caption{(Color online)  Phase diagram obtained from plotting the discontinuity $\Delta n$, in the disorder strength $W$ interaction strength $V$ plane, of the Aubry-Andr{\'e} model for system size $L=16$ and energy density $\varepsilon = 1$ ($V'=0$).}
\label{fig:phasediagramAAH}
\end{figure}

We verify that the occupation discontinuity is also obtained in the interacting Aubry-Andr{\'e} model and use it to map out its phase diagram.
This model is given by the Hamiltonian~\eqref{eq:ham} with the quasirandom on-site disorder potential
\begin{equation}
	\epsilon_i = W \cos(2\pi \beta i + \phi), 
\end{equation}
where $\beta = \frac{\sqrt{5}-1}{2}$ and $\phi$ is a random phase. 
Experimental studies of many-body localization in the same model, albeit with spinful fermions and with onsite interactions, were reported in Refs.~\onlinecite{Schreiber2015,Bordia2016}.
Prior to that, a realization of the Aubry-Andr{\'e} model was used to investigate Anderson localization \cite{Roati2008} and interacting systems at low energy densities 
\cite{DErrico2014} with cold atoms,
triggering many theoretical studies of its low-energy properties  in the presence of interactions, for both bosons (see, e.g., \cite{Roux2008,Roscilde2008}) and 
fermions (see, e.g, \cite{Mastropietro2015}).
In the absence of interactions, the Aubry-Andr{\'e} model has a metal-insulator transition at $W_c = 2$ \cite{Aubry1980} and interactions increase the critical disorder strength a little 
bit, to about $W_c = 2.5$ depending on the interaction strength \cite{Iyer2013}.

In Fig.~\ref{fig:occupationsAAH}, we plot the occupation spectrum for different disorder strengths (see \cite{Iyer2013,Modak2015,Naldesi2016} for other studies of many-body localization in this system).
As in the random model, the occupation spectrum has a clear discontinuity in the many-body localized phase, while in the ergodic phase, it is absent (up to $1/L$ corrections).
From the discontinuity at different disorder strengths $W$ and  interaction strengths $V$ we obtain the phase diagram given in Fig.~\ref{fig:phasediagramAAH}.
This phase diagram is consistent with what is already known about this model \cite{Iyer2013,Naldesi2016,Schreiber2015}, i.e., interactions lead to a slight increase of the 
delocalized phase.
The example of spinless fermions with nearest-neighbor interactions in the Aubry-Andr{\'e} model demonstrates that our main findings, i.e., the existence of a discontinuity in the occupation spectrum, is not restricted to the model of spinless fermions in a random lattice.

\section{Natural orbitals as an optimum measure for the single-particle content of quasiparticles}
\begin{figure}[tb]
\includegraphics[width=1\columnwidth]{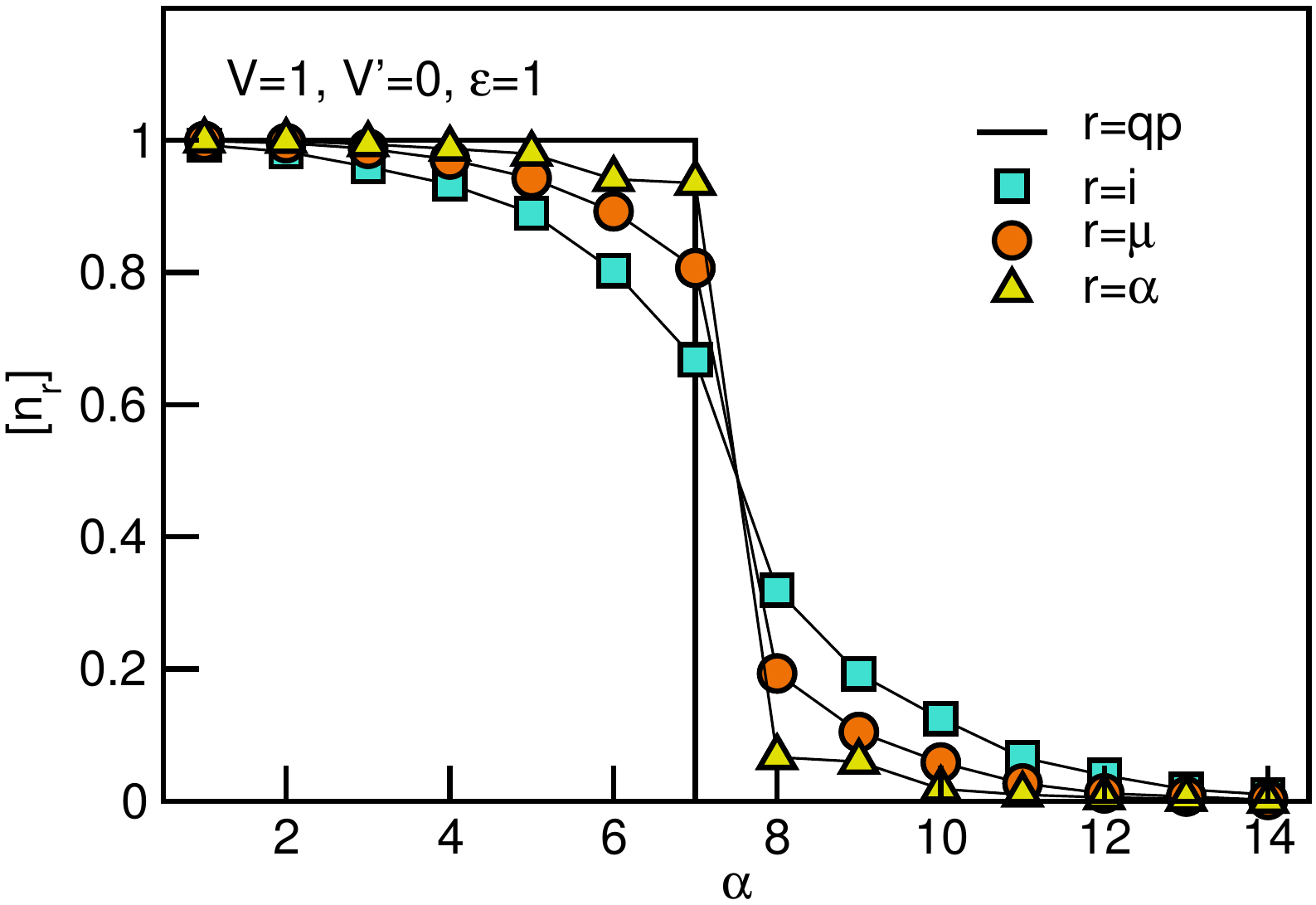}
\caption{(Color online) Spinless fermions with nearest-neighbor repulsion and random onsite disorder: comparison of three different occupations $[n_r]$, the real-space occupation $n_i$, the occupation of Anderson orbitals $n_\mu$ and that natural orbital occupations $n_\alpha$, ordered in descending order. The solid line is the
(step-like) distribution function of the quasiparticles. Here $V=1.0$, $W=6.0$ and $L=14$, for many-body states from the middle of the spectrum. We  averaged over 64 disorder realizations. }
\label{fig:nmu_vs_nalpha}
\end{figure}

In this section we discuss the usefulness of the natural orbitals as a basis for a quasiparticle expansion and as an optimum measure for the single-particle content
 of the quasiparticles.
The motivation for this is as follows: in the bulk of the paper we have, based on the numerical analysis, demonstrated how the occupations of the natural orbitals characterize the full many-body localized phase.
As far as formal operator expansions for the quasiparticles are concerned, one may worry, however, that the perturbative expression~\eqref{eq:qpcoeffs} is only useful when the effect of interactions is weak, that is, when the natural orbitals are very close to the Anderson orbitals. 
In this limit, the expectation value $n_\mu = \langle n |\hat{n}_\mu | n\rangle$ of the densities of the Anderson orbitals should also be a good approximation of the quasiparticle occupations.
In Fig.~\ref{fig:nmu_vs_nalpha}, we compare this density $n_\mu$ with both the onsite densities $n_i = \langle n|\hat{n}_i|n\rangle$ and the natural orbital occupations $n_\alpha$; in all cases, we have ordered them in descending order.
The plot demonstrates that the natural orbitals give the best approximation to the quasiparticle occupations since they are the closest to integers.
This is a central result of our study.

Given this and the fact that the whole many-body localized phase has well defined quasiparticles, also at strong interactions, one may be tempted to express the quasiparticles in terms of the natural orbitals. 
That is, instead of Eq.~\eqref{eq:lbits2}, one writes
\begin{equation}
        \hat{{n}}_\alpha^{(\rm qp)} \simeq  \hat{n}_\alpha + \sum_{(\beta\beta^\prime) \neq (\gamma\gamma^\prime)} \tilde B^{\alpha,(n)}_{\beta\beta^\prime\gamma\gamma^\prime} c^\dagger_\beta c^\dagger_{\beta^\prime}c_{\gamma}c_{\gamma^\prime} + \ldots\,.
\label{eq:lbit_exp_opdm_p}
\end{equation} 
This form is still consistent with the expectation that $\hat{{n}}_\alpha^{(\rm qp)} \to \hat n_{\mu}$ in the noninteracting limit since in that case, the diagonalization of the OPDM just gives the single-particle energy eigenstates.
Equation~\eqref{eq:lbit_exp_opdm_p} will generally not be true at the operator level, since for the one-particle sector, $\hat{n}^\alpha$ may not generally commute with the single-particle Hamiltonian.
However, for the $N$-particle sector to which the eigenstate $|n\rangle$ belongs to, and when evaluated in the state $|n\rangle$, this expression is likely to be accurate (this is the reason for the $\simeq$ symbol).
The slight disadvantage of the form Eq.~\eqref{eq:lbit_exp_opdm_p} is that the actual operator expansion of the integrals of motion now varies from many-body eigenstate to many-body eigenstate, although as shown in Fig.~\ref{fig:nbeta} this variation is not too severe.

As a consequence of Eq.~\eqref{eq:lbit_exp_opdm_p}, the quasiparticle creation operator is related to the natural-orbital operators as
\begin{equation}
        c^{({\rm qp})\dagger}_\alpha = c_\alpha^\dagger + \sum_{\beta\gamma\delta} f^\alpha_{\beta\gamma\delta} c^\dagger_\beta c_\gamma^\dagger c_\delta + \ldots,
		\label{eq:qp_expansion}
\end{equation}
This is reminiscent of the relation of quasiparticles in terms of plane waves in Fermi-liquid theory (see, for example, Eq.~(8) in Ref.~\onlinecite{Varma:2002wl} and references within).
Plugging Eq.~\eqref{eq:qp_expansion} into Eq.~\eqref{eq:lbit_exp_opdm_p} and evaluating the result in the state $|n\rangle$, we obtain, to first order in $f$,
\begin{equation}
	n_\alpha = n_\alpha^{(\rm qp)}\left[1-2\text{Re}\sum_\beta (f^\alpha_{\alpha\beta\beta}-f^\alpha_{\beta\alpha\beta})n^{(\rm qp)}_\beta \right].
\end{equation}
The advantage of this relation over the similar relation in terms of the Anderson orbitals~\eqref{eq:qpcoeffs} is that here it is explicit that any deviation from the ideal step function in the occupations comes only from particle-hole dressing of bare fermions in the quasiparticle expansion.
Furthermore, since this expression is purely in terms of the natural orbitals, it provides a reasonable approximation at any interaction strength at which quasiparticles with nonzero single-particle content emerge.

It remains as an open question to mathematically derive a nonperturbative relation between natural orbitals and integrals of motions/quasiparticles. 
Moreover, our results suggest that in any many-body eigenstate, the OPDM eigenbasis is the single-particle basis that best approximates the quasiparticles. 
Another interesting extension would be to ask whether there is a global single-particle basis (perhaps obtained from suitable linear combinations of natural orbitals from different states) that  provides a further improvement along these lines.

\section{Summary and outlook}
\label{sec:summary}
In this work, we provided a comprehensive discussion of the properties of the one-particle  density matrix in one-dimensional systems of 
interacting fermions in the presence of disorder. We showed that the eigenstates of the OPDM, which form complete sets of single-particle states, computed in individual many-body eigenstates,
are delocalized in the ergodic phase and localized in the MBL phase. The eigenvalues, after a suitable reordering, unveil the 
Fock-space structure of MBL many-body eigenstates: they are weakly dressed Slater determinants \cite{Basko2006} since most eigenvalues are close to either one or zero and the occupation spectrum 
has a discontinuity at an emergent Fermi edge in the MBL phase. This suggests a close analogy of the MBL phase to a zero-temperature Fermi liquid and is consistent
with the existence of localized quasiparticles in the MBL phase. We relate these findings to the local integrals of motion  introduced to describe the phenomenology of the
MBL phase \cite{Vosk2013,Serbyn2013,Huse2014,Imbrie2016,Imbrie2016a}. We argue that the OPDM eigenstates provide a good (if not the best) approximation to the single-particle content of the local integrals of motion.
In the ergodic phase, the occupation spectrum is thermal in agreement with the eigenstate thermalization hypothesis and the 
occupation-spectrum discontinuity vanishes exponentially or faster as system size increases.
Computing the occupation-spectrum discontinuity as a function of model parameters thus provides a sharp measure of the phase diagram.

We further discussed the statistical properties of both natural orbitals and of the occupation spectrum. Interestingly, there are increasing fluctuations
in the occupation-spectrum discontinuity as the transition is approached. We demonstrate that there is a correlation between states with a small
occupation-spectrum discontinuity and a large entanglement entropy. Similar to the conclusions drawn in other works, these fluctuations increase as the 
transition is approached and these fluctuations are thus indications of the breakdown of localization. 

While most of our analysis focussed on the paradigmatic model of spinless fermions with nearest-neighbor repulsive interactions, we also 
show that computing the occupation-spectrum discontinuity correctly resolves the ergodic and MBL phases in a model of interacting spinless
fermions in the Aubry-Andr{\'e} model as well.

We conclude by pointing out several interesting applications of our single-particle framework to other models that are of relevance in the 
theory of many-body localization.
An interesting case for the OPDM analysis to be applied to is the Fermi-Hubbard model \cite{Mondaini2015,BarLev2016,Sierant2016,Prelovsek2016h}, 
where disorder has been introduced by either disorder that couples to the charge or in the interactions. In the latter case, 
there is not Anderson-insulator in the non-interacting limit and hence it would be very interesting to see whether the 
natural orbitals become localized in spite of the single-particle eigenstates all being delocalized. 
In the former case, recent work has raised doubts on the existence of localization and nonergodic dynamics in the spin sector \cite{Prelovsek2016h}, in which
the chosen type of disorder preserves SU(2) symmetry, which has been argued to protect the system from localizing \cite{Potter2016}.

A crucial question that is not fully understood yet is to which degree signatures of MBL can still be seen in open quantum systems (for some work in this direction, see
\cite{Johri2015,Nandkishore2014,Fischer2016, Prelovsek2016, Znidaric2016, Hyatt2016,Chandran:2016ji}).
That would be the generic situation in condensed matter systems, where a coupling to phonons is usually unavoidable. In the case of the Anderson insulator,
it is known that phonons turn the system into a bad metal \cite{Kramer1993,Evers2008,Basko2006}.
The notion of localized quasiparticles provides a framework to address this question. Can the lifetimes $\tau_{\rm qp}$ of the quasiparticles in the presence of a bath
be long enough as to allow MBL physics to survive at least on times $t< \tau_{\rm qp}$?  The analysis of the OPDM in an open quantum system and the
evolution of the discontinuity as a function of bath coupling would be worth looking at.

{\it Acknowledgment.} We thank  D. Abanin, E. Altman, D. Logan, F. Pollmann, and L. Rademaker  for insightful discussions. 
This work was supported by the ERC Starting Grant No. 679722, and in part by the National Science Foundation under Grant No. NSF PHY11-25915. 
\bibliography{references}

\end{document}